\documentclass[twocolumn, amssymb, amsmath, aps, superscriptaddress, showpacs, footinbib,  prb]{revtex4}
\pdfoutput=1
\usepackage{graphicx}

\newcommand{\eff}{\text{eff}}
\newcommand{\AFM}{\text{AFM}}

\begin{document}

\title{Structural distortion and frustrated magnetic interactions\\ in the layered copper oxychloride (CuCl)LaNb$_2$O$_7$}
\author{Alexander A. Tsirlin}
\email{altsirlin@gmail.com}
\affiliation{Max Planck Institute for Chemical Physics of Solids, N\"{o}thnitzer
Str. 40, 01187 Dresden, Germany}
\affiliation{Department of Chemistry, Moscow State University, 119992 Moscow, Russia}
\author{Helge Rosner}
\email{Helge.Rosner@cpfs.mpg.de}
\affiliation{Max Planck Institute for Chemical Physics of Solids, N\"{o}thnitzer
Str. 40, 01187 Dresden, Germany}
\date{\today}

\begin{abstract}
We present a computational study of the layered copper oxychloride (CuCl)LaNb$_2$O$_7$ that has been recently proposed as a spin-1/2 frustrated square lattice compound. Our results evidence an orbitally degenerate ground state for the reported tetragonal crystal structure and reveal a Jahn-Teller-type structural distortion. This distortion heavily changes the local environment of copper -- CuO$_2$Cl$_2$ plaquettes are formed instead of CuO$_2$Cl$_4$ octahedra -- and restores the single-orbital scenario typical for copper oxides and oxyhalides. The calculated distortion is consistent with the available diffraction data and the experimental results on the electric field gradients for the Cu and Cl sites. The band structure suggests a complex three-dimensional spin model with the interactions up to the fourth neighbors. Despite the layered structure of (CuCl)LaNb$_2$O$_7$, the spin system has pronounced one-dimensional features. Yet, sizable interchain interactions lead to the strong frustration and likely cause the spin-gap behavior.  Computational estimates of individual exchange couplings are in qualitative agreement with the experimental data. 
\end{abstract}

\maketitle

\section{Introduction}
Orbital ordering is one of the unusual and attractive phenomena in solid state physics. Specific occupation of orbitals plays a major role in electronic properties of numerous transition metal compounds. Orbital ordering is generally accompanied by the Jahn-Teller effect, a structural distortion that lifts the orbital degeneracy.\cite{kugel1982} The Jahn-Teller distortion mainly affects the local environment of the transition metal cation, while the crystallographic unit cell is only slightly changed or even unchanged at all. The latter issue leads to certain difficulties in the structure analysis, since the distortion causes minor alterations of the diffraction patterns. Nevertheless, modern experimental techniques of high-resolution neutron and synchrotron x-ray diffraction are usually able to resolve weak structural changes associated with the orbital ordering (see, e.g., Refs.~\onlinecite{smvo3,dyvo3,sr2vo4,bimno3}).

Divalent copper is probably the most known transition metal cation subjected to the Jahn-Teller effect. In a regular octahedral environment, the electronic configuration $d^9$ of Cu$^{+2}$ leads to the orbital degeneracy that is usually lifted by an extremely strong tetragonal distortion. This distortion reduces the coordination number of copper and yields CuO$_4$ plaquettes typical for the structures of Cu$^{+2}$-containing oxides.\cite{wells} High-symmetry structures (e.g., perovskite structure) may constrain the Jahn-Teller distortion from the plaquettes formation. Nevertheless, a notable tetragonal distortion takes place, the orbital ordering is established, and peculiar electronic properties emerge. Thus, perovskite-type copper fluorides KCuF$_3$ and K$_2$CuF$_4$ are known as first examples of the cooperative Jahn-Teller distortion, giving rise to ferromagnetic (FM) interactions induced by the specific orbital ordering.\cite{kugel1982,wells,khomskii1973} 

\begin{figure}
\includegraphics{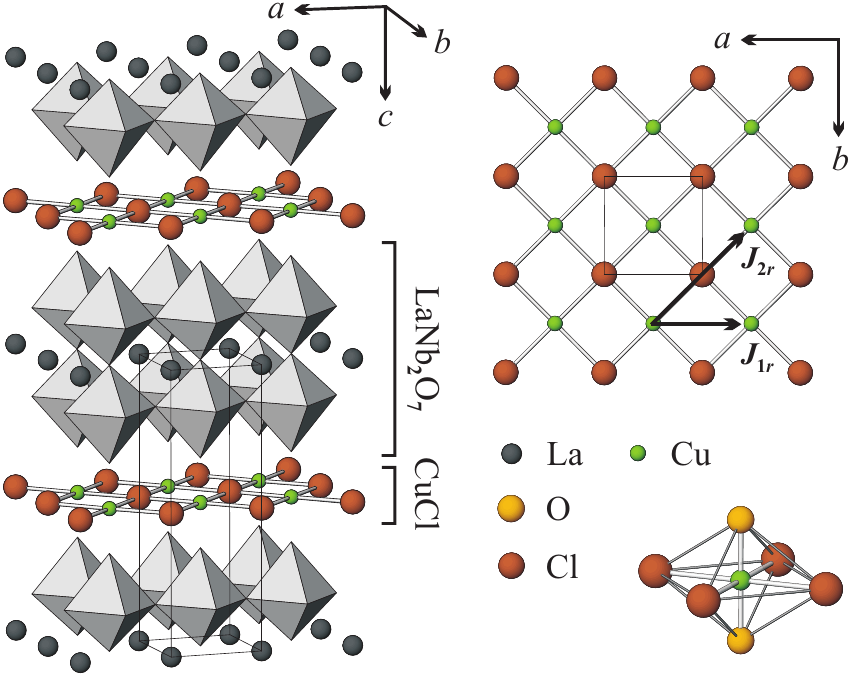}
\caption{\label{fig_structure}(Color online)
Regular (tetragonal) crystal structure of (CuCl)LaNb$_2$O$_7$: stacking of perovskite-type [LaNb$_2$O$_7$] blocks and rocksalt-type [CuCl] layers (left panel), the single [CuCl] layer (upper right panel), and the CuO$_2$Cl$_4$ squeezed octahedron (bottom right panel). The regular structure of the [CuCl] layer gives rise to the frustrated square lattice spin model with the competing nearest-neighbor and next-nearest-neighbor couplings $J_{1r}$ and $J_{2r}$, respectively.
}
\end{figure}

Despite the large magnitude in the Jahn-Teller distortion in copper compounds, the identification of the distortion pattern may be quite problematic. This is the case for layered copper oxyhalides (CuX)LaM$_2$O$_7$ with X = Cl, Br and M = Nb, Ta. These compounds have composite structures built by [LaM$_2$O$_7$] perovskite-type blocks and [CuX] rocksalt-type layers (see Fig.~\ref{fig_structure}).\cite{kodenkandath1999,kodenkandath2001} In the following, we will focus on one of these compounds, (CuCl)LaNb$_2$O$_7$, that recently drew attention due to its unusual and puzzling magnetic properties.\cite{kageyama2005,kageyama2005-2}

Initially, the crystal structure of (CuCl)LaNb$_2$O$_7$ was refined in the tetragonal space group $P4/mmm$ with the Cl atom located in the special $1b$ position $(0,0,\frac12)$.\cite{kodenkandath1999} In this structure (further referred as regular), copper has a squeezed octahedral coordination with two short Cu--O bonds [$d$(Cu--O) = 1.97 \r A] and four long Cu--Cl bonds [$d$(Cu--Cl) = 2.74 \r A], see Fig.~\ref{fig_structure}. This type of local environment is quite unusual for copper oxychlorides: normally, these compounds reveal the Jahn-Teller effect and the square-planar CuCl$_4$ or CuO$_2$Cl$_2$ coordination with $d$(Cu--Cl)$=2.3-2.4$ \r A.\cite{wells} Moreover, the refined Debye-Waller factor for the Cl atom is extremely high ($U_{\text{iso}}=0.13$ \r A$^2$) suggesting a shift of Cl away from the $1b$ position.\cite{kodenkandath1999} Caruntu \textit{et al}.\cite{caruntu2002} proposed a new structural model with the Cl atoms randomly occupying one quarter of $4m$ sites $(x,0,\frac12)$ with $x=0.136$. This model yields two short (about 2.4 \r A) and two long (about 3.15~\r A) \mbox{Cu--Cl} distances consistent with the crystal chemistry of copper oxychlorides.\cite{wells} Caruntu \textit{et al}.\cite{caruntu2002} tentatively ascribed the distortion to the Jahn-Teller effect of Cu$^{+2}$. However, they failed to observe any superstructure reflections that could arise due to the cooperative Jahn-Teller distortion, similar to KCuF$_3$ and K$_2$CuF$_4$.\cite{kugel1982,wells,khomskii1973}

The structure of (CuCl)LaNb$_2$O$_7$ was further studied by means of nuclear magnetic resonance (NMR) and nuclear quadrupole resonance (NQR) measurements.\cite{yoshida2007} The spectra evidenced the lack of the tetragonal symmetry and revealed single sites of Cu, Cl, and La atoms, hence suggesting the ordering of the Cl atoms, at least on the local scale. Additionally, Yoshida \textit{et al}.\cite{yoshida2007} performed electron diffraction studies and found superstructure reflections that unambiguously confirmed the ordering of the Cl atoms and the resulting cooperative distortion of the copper polyhedra. However, the specific ordering pattern remains unclear. To refine the superstructure, one has to use the superstructure reflections that are absent in x-ray and neutron diffraction patterns.\cite{kodenkandath1999,caruntu2002,oba2007} The respective reflections are revealed by electron diffraction, but their intensities are strongly affected by multiple scattering and hence unsuitable for the refinement.\cite{foot1} 

The magnetic properties of (CuCl)LaNb$_2$O$_7$ are unusual and lack a clear microscopic interpretation. The studies reveal a spin gap behavior and a singlet ground state that are incompatible with the frustrated square lattice ($J_{1r}-J_{2r}$) model, as proposed by the regular crystal structure in Fig.~\ref{fig_structure}.\cite{kageyama2005,kageyama2005-2,kitada2007} Inelastic neutron scattering (INS) data are even more puzzling and suggest long-range interactions (between Cu atoms separated for about 9~\r A) to be relevant for (CuCl)LaNb$_2$O$_7$.\cite{kageyama2005} Such an unusual magnetic behavior could be caused by a non-trivial orbital state of copper and intricate superexchange pathways, emerging due to a specific cooperative distortion. Thus, unraveling this distortion is a key to understand the magnetic properties of (CuCl)LaNb$_2$O$_7$.

Presently, there are two reports that consider possible ordering patterns of the Cl atoms and attempt to relate these patterns to the exchange interactions in (CuCl)LaNb$_2$O$_7$. Whangbo and Dai\cite{whangbo2006} used extended H\"uckel calculations to study qualitatively exchange couplings in several ordered structures. They proposed a model of ring clusters with a number of inequivalent Cu and Cl sites that is in contradiction to the NMR and NQR results.\cite{yoshida2007} Yoshida \textit{et~al}.\cite{yoshida2007} employed an even more empirical approach and searched for an ordering pattern with "explicit"\ copper dimers (i.e., two Cu atoms connected by the double bridge of short Cu--Cl bonds). Such dimers can probably account for the spin-gap behavior, but they are unable to explain the long-range interactions, as proposed by the INS data. Neither of the two studies considered the underlying reasons for the structural distortion or proposed numerical estimates of the exchange couplings.

The physics of (CuCl)LaNb$_2$O$_7$ is quite complex and includes both structural (cooperative distortion) and magnetic (exchange interactions) aspects. Unfortunately, experimental techniques up to now fail to determine the superstructure, hence complicating the analysis of the exchange couplings. The latter is non-trivial itself due to the presence of long-range interactions. In this work, we attack the problem by using full-potential band structure calculations based on the density functional theory (DFT), since computational methods are known as an effective tool for solving complex problems in solid state physics. The main advantage of the computational approach is the possibility to estimate numerically the relevant parameters such as the total energy and the exchange couplings. The evaluation of the total energy is crucial for resolving the structural problems in case conventional diffraction techniques experience difficulties in a precise structure determination. For example, the local density approximation (LDA) level of DFT was successfully applied for studying the structures of lithium monoboride,\cite{lib} while proper treatment of correlation effects allowed to derive Jahn-Teller distortion and to explain the resulting FM interactions in transition-metal fluorides KCuF$_3$ (Refs.~\onlinecite{kcuf3} and~\onlinecite{kcuf3-2}) and Cs$_2$AgF$_4$ (Ref.~\onlinecite{cs2agf4}). Quite recently, band structure calculations predicted ferroelectric distortions in unconventional manganite multiferroics\cite{tbmn2o5-computation,homno3} followed by the experimental confirmation.\cite{tbmn2o5-exp}

The outline of the paper is as follows. We start with the discussion of methodological aspects in Sec.~\ref{methods}. In Sec.~\ref{tetragonal}, we present the computational results for the regular structure of (CuCl)LaNb$_2$O$_7$. We find orbital degeneracy and show that the regular structure is in qualitative contradiction to the experimental magnetic behavior. In Sec.~\ref{distorted}, we derive the relevant structural distortion, evaluate exchange integrals, and propose a realistic spin model of (CuCl)LaNb$_2$O$_7$. In Sec.~\ref{discussion}, we compare our results to the experimental findings, discuss the resulting spin model, and demonstrate that our scenario provides consistent interpretation of the experimental data. We also give an outlook for further experimental studies and conclude the section with a short summary.

\section{Methods}
\label{methods}
Scalar-relativistic band structure calculations were performed using a full-potential local-orbital scheme (FPLO, versions 7.00-27 and 8.50-32)\cite{fplo} and the exchange-correlation potential by Perdew and Wang.\cite{perdew} LDA calculations were done for the smallest possible unit cells employing the actual symmetry of the atomic positions, i.e., a 12-atom tetragonal unit cell for the regular structure ($a\times a\times c$, space group $P4/mmm$) and a 24-atom orthorhombic unit cell for the distorted structure ($a\times 2a\times c$, space group $Pbmm$). For the local spin density approximation (LSDA)+$U$ calculations with different types of spin ordering, a number of monoclinic 48- and 72-atom supercells were used (space groups $Pm$ or $P2/m$ with $c$ being the monoclinic axis, see Sec.~\ref{distorted} and Fig.~\ref{fig_scells} for details). The structure optimization was carried out in the 48-atom $2a\times 2a\times c$ supercell, as suggested by the electron diffraction data.\cite{yoshida2007} The triclinic ($P1$) symmetry of this supercell allowed for free relaxation of all 48 atoms. Different unit cells employed different $k$ meshes with at least 72, 192, 730, and 2925 $k$ points in the irreducible part of the first Brillouin zone for the 72-, 48-, 24-, and 12-atom cells, respectively. The convergence with respect to the $k$ mesh was carefully checked.

Two sets of structural information are available for (CuCl)LaNb$_2$O$_7$. In Ref.~\onlinecite{kodenkandath1999}, the structure was refined from the x-ray data, while in Ref.~\onlinecite{caruntu2002} the neutron data were used. The latter results are claimed to be more accurate, because neutrons are more sensitive to the positions of light atoms.\cite{caruntu2002} Additionally, the interatomic distances yield more reasonable formal valences, as calculated with the empirical bond valence sum rules.\cite{caruntu2002} Yet, the authors of Ref.~\onlinecite{caruntu2002} note that the resulting Cu--O distance (1.84 \r A) is unusually short as compared to other copper oxides and oxychlorides.\cite{wells} Keeping in mind this ambiguity, we performed calculations for both sets of the atomic coordinates. The results for the stabilization energies and the exchange couplings are similar within $5-10$\%, which is likely comparable to the actual accuracy of our calculations. This outcome is quite natural, because the difference between the two structures deals with the atomic positions along the $c$ axis (the positions in the $ab$ plane are fixed by the symmetry), whereas the main physics takes place in the $ab$ plane. Note, however, that the structure refined from the neutron data has lower energy, though both structures are not fully relaxed and reveal sizable forces ($0.3-0.4$ eV/a.u.) along the $c$ axis for Nb and O atoms. The forces on the respective atoms have opposite signs, hence the relaxed atomic positions lie between those suggested by x-rays and neutrons.

The general computational procedure for both regular and distorted structures is as follows. We start with LDA calculations, focus on the states close to the Fermi level and construct an effective model for these states. The model yields the estimates for all the antiferromagnetic (AFM) couplings in the systems under investigation, hence facilitating the choice of relevant interactions. We also calculate Wannier functions in order to get a direct, graphical representation of the relevant orbitals. The Wannier functions are computed in a maximally localized fashion\cite{wannier} using the internal procedure of FPLO8.50. Then we proceed to LSDA+$U$ calculations and get independent estimates of the leading exchange couplings. Additionally, we use LSDA+$U$ for the structure optimization, because LDA is known to conceal Jahn-Teller distortions in transition metal compounds due to improper treatment of the correlation effects.\cite{kcuf3,cs2agf4} 

In the LDA band structure, the states close to the Fermi level are relevant for the possible orbital degeneracy and for the magnetic interactions. The respective bands are analyzed within a tight-binding (TB) model, and the resulting hoppings ($t$) are introduced to an extended Hubbard model with the effective on-site Coulomb repulsion $U_{\eff}$. For both regular and distorted structures, the strongly correlated limit $t\ll U_{\eff}$ is realized, hence the Hubbard model can be reduced to the Heisenberg model for the low-lying excitations under the assumption of the half-filling. The AFM contributions to the exchange integrals are estimated as $J_i^{\AFM}=4t_i^2/U_{\eff}$, and all the possible superexchange pathways are analyzed. The latter feature is especially important, since long-range interactions are expected in (CuCl)LaNb$_2$O$_7$.\cite{kageyama2005} The precise value of the $U_{\eff}$ parameter for copper oxyhalides is unknown, although a recent study of CuCl$_2$ and CuCl$_2\cdot 2$H$_2$O suggests $U_{\eff}=4$~eV.\cite{cucl2,foot2} In the present work, we employ the representative value $U_{\eff}=4.5$ eV, as commonly used for copper oxides (see, e.g., Refs.~\onlinecite{sr2cup2o8,bi2cuo4,licuvo4,cusb2o6}). Note that the values of $J_i^{\AFM}$ merely scale with $U_{\eff}$, and the ratios of $J_i^{\AFM}$, which are the relevant quantity for the ground state, are not affected by the ambiguity of the choice of $U_{\eff}$.

The LDA results can be effectively used to construct models on top of DFT. However, it is also helpful to account for the correlation effects in the course of the self-consistent calculations. This opportunity is provided by the LSDA+$U$ method that treats the correlation effects in the mean-field approximation. LSDA+$U$ is known to yield reasonable estimates of different quantities [energy gaps, electric field gradients (EFGs), and exchange integrals] in strongly correlated electronic systems and, in particular, to treat correctly the cooperative Jahn-Teller distortions.\cite{kcuf3,cs2agf4} The problematic point of the LSDA+$U$ approach deals with the choice of the $U_d$ and $J_d$ values that account for the on-site Coulomb repulsion and exchange, respectively. The optimal values of these parameters are known to depend on the particular computational method and on the objective quantities as well as on the specific chemical and structural features of the compound under investigation.\cite{kcuf3,sr2cup2o8,bi2cuo4} Unfortunately, there are no established LSDA+$U$ parameters for copper oxychlorides, although the recent application of $U_d=6.0-8.5$~eV to CuCl$_2$ and CuCl$_2\cdot 2$H$_2$O can be used as a reference.\cite{cucl2,foot2} In our calculations, we employ several representative and physically reasonable values of $U_d$ over a wide range, namely, 3.5, 5.5, and 7.5 eV.\cite{foot4} Below, we will show that all these values lead to qualitatively similar behavior, although the details of the microscopic scenario depend on the $U_d$ value used. The exchange parameter $J_d$ is fixed at 1~eV, since it is known to have minor influence on the results. 

One should keep in mind that the LSDA+$U$ Coulomb repulsion parameter $U_d$ is usually \textit{different} from the effective on-site Coulomb repulsion $U_{\eff}$, as employed in the model analysis above. The latter potential corresponds to the antibonding $pd\sigma^*$ LDA bands with the contributions from Cu, Cl, and O orbitals (see Fig.~\ref{fig_dost}), while $U_d$ is applied to the Cu $3d$ orbitals \textit{only}. In the framework of the one-band Hubbard model, the antiferromagnetic superexchange couplings should scale as $1/U_{\eff}$, but this scaling is not necessarily applicable to the LSDA+$U$ results for the total exchange and to their dependence on $U_d$.

\section{Regular structure}
\label{tetragonal}
\begin{figure}
\includegraphics[scale=0.95]{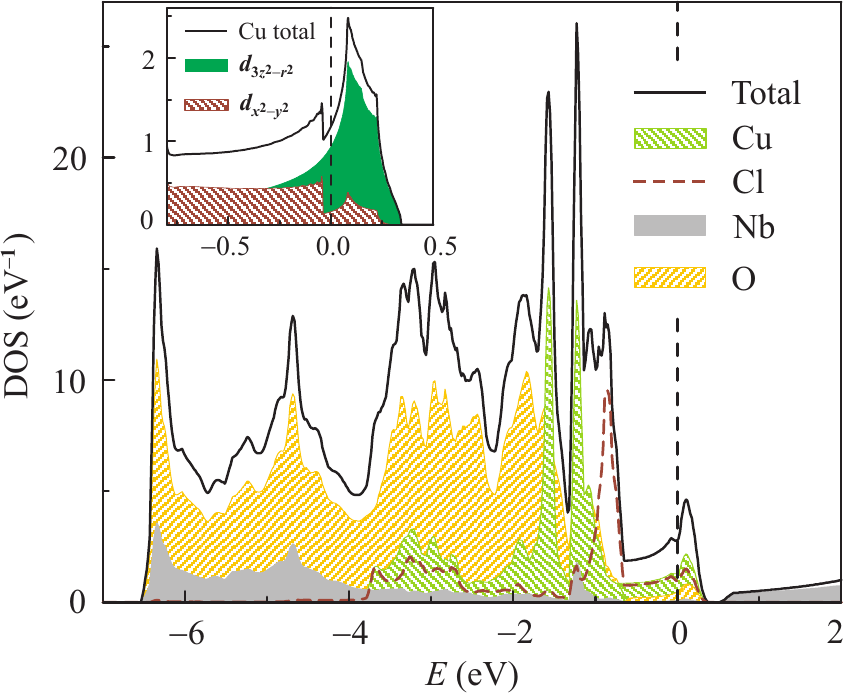}
\caption{\label{fig_dost}(Color online)
LDA density of states (DOS) for the regular (tetragonal) structure of (CuCl)LaNb$_2$O$_7$. The Fermi level is at zero energy, the states close to the Fermi level are formed by Cu, Cl, and O orbitals. The inset shows a part of the orbital resolved DOS for copper with two orbitals ($d_{3z^2-r^2}$ and $d_{x^2-y^2}$) contributing to the upper valence bands.
}
\end{figure}

The LDA density of states (DOS) for the regular structure of (CuCl)LaNb$_2$O$_7$ is shown in Fig.~\ref{fig_dost}. The energy spectrum is similar to that in other copper oxychlorides.\cite{cute2252,cute45124} The low-lying valence bands are predominantly formed by oxygen orbitals, while at higher energies copper and chlorine states become prominent. The states near the Fermi are composed of copper orbitals, with sizable contributions from oxygen and chlorine. Conduction bands are formed by niobium and oxygen orbitals. The spectrum is consistent with the intuitive ionic picture, suggesting the oxidation states of +2 and +5 for Cu ($3d^9$) and Nb ($4d^0$), respectively. Thus, peculiar magnetic properties originate from the [CuCl] layers, while the Nb-containing [LaNb$_2$O$_7$] blocks are non-magnetic and insulating. 

The energy spectrum is metallic in contradiction to the observed green color of (CuCl)LaNb$_2$O$_7$.\cite{kodenkandath1999} The unrealistic metallicity is a typical failure of LDA due to the underestimate of the correlation effects. The application of LSDA+$U$ readily restores an energy gap of $1.0-1.2$ eV for both the regular and distorted structures. Such a gap is still insufficient to account for the green color of the compound. However, one should be aware that the straightforward mapping of either the LDA or the LSDA+$U$ single-particle spectrum onto the optical properties may be misleading. Many-body effects are sometimes crucial for proper evaluation of the excitation energies, as recently shown for lithium niobate, LiNbO$_3$.\cite{linbo3} We believe that the optical properties of (CuCl)LaNb$_2$O$_7$ require a careful study on the many-body level of theory, and such a study lies beyond the scope of the present work. In the following, we restrict ourselves to the discussion of ground-state properties (optimized crystal structure) and low-lying excitations (magnetic interactions). Both problems deal solely with the valence bands, while the conduction bands lie $1-2$~eV higher in energy hence bearing little influence on the objectives of our work.

\begin{figure}
\includegraphics[scale=0.9]{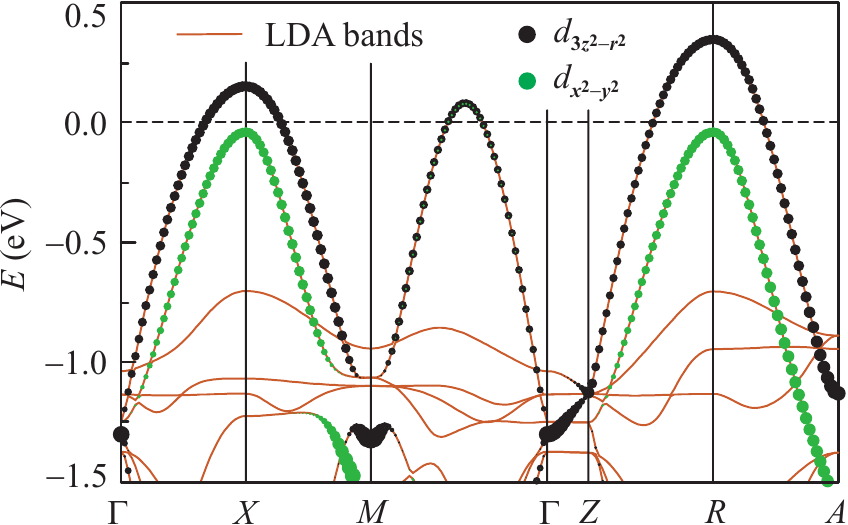}
\caption{\label{fig_bandt}(Color online)
LDA bands for the regular (tetragonal) crystal structure of (CuCl)LaNb$_2$O$_7$. Lines show the bands, while dots indicate the contributions of the $d_{3z^2-r^2}$ and $d_{x^2-y^2}$ orbitals. The Fermi level is at zero energy. The notation of $k$ points is as follows: $\Gamma(0,0,0)$, $X(0.5,0,0)$, $M(0.5,0.5,0)$, $Z(0,0,0.5)$, $R(0.5,0.5,0.5)$, and $A(0.5,0,0.5)$, where the coordinates are given along the $k_x$, $k_y$, and $k_z$ axes in units of the respective reciprocal lattice parameters $2\pi/a$, $2\pi/a$, and $2\pi/c$.
}
\end{figure}

The orbital resolved DOS shows that the bands near the Fermi level are mainly formed by Cu $d_{3z^2-r^2}$ orbital with a contribution from the $d_{x^2-y^2}$ orbital (see the inset of Fig.~\ref{fig_dost}). The same picture emerges from the band structure plot. There are two distinct copper bands above $-1$ eV, and these bands are formed by the $d_{3z^2-r^2}$ and $d_{x^2-y^2}$ orbitals (Fig.~\ref{fig_bandt}). Crystal field considerations suggest that the orbital with the highest energy should point to the nearest ligands. Indeed, the $d_{3z^2-r^2}$ orbital points to oxygen atoms with short Cu--O separations of about 1.85 \r A. The in-plane Cu--Cl separations are much longer (2.74 \r A), but the chlorine orbitals are spatially more extended as compared to the oxygen orbitals. Thus, both $e_g$ orbitals of copper have similar energies, and the LDA ground state is orbitally degenerate. Correlations induce an orbital ordering within the regular structure, but -- as we will show below -- such orbital ordering is unable to account for the magnetic properties of (CuCl)LaNb$_2$O$_7$.

The $d_{3z^2-r^2}$ band can be analyzed within a simple TB model, including nearest-neighbor ($t_{1r}$) and next-nearest-neighbor ($t_{2r}$) hoppings (the subscript $r$ refers to the regular structure). We find $t_{1r}^{3z^2-r^2}=-3$ meV and $t_{2r}^{3z^2-r^2}=183$ meV. The fitting of the $d_{x^2-y^2}$ band is more complicated due to its strong hybridization with the chlorine orbitals. However, one can roughly estimate $t_{1r}^{x^2-y^2}=-44$~meV and $t_{2r}^{x^2-y^2}=172$~meV from the $\Gamma-X-M$ part of the band structure. Thus, both orbitals yield the $t_{2r}\gg t_{1r}$ scenario and suggest the strong AFM interaction $J_{2r}^{\AFM}$ of about $300-350$~K. 

Electron correlations are able to lift the orbital degeneracy without involving a structural distortion, as recently shown for CuSb$_2$O$_6$.\cite{cusb2o6} In this compound, the LDA band structure reveals similar contributions of the $d_{3z^2-r^2}$ and $d_{x^2-y^2}$ orbitals to the states near the Fermi level. The introduction of electron correlations within LSDA+$U$ enables to stabilize the half-filling for either of the orbitals, but the $d_{3z^2-r^2}$ -type orbital ordering has lower energy and yields the realistic physical picture. The case of (CuCl)LaNb$_2$O$_7$ is different. LSDA+$U$ leads to the half-filling of the $d_{3z^2-r^2}$ orbitals (the state with the $d_{x^2-y^2}$-type ordering can not be stabilized), but the resulting scenario contradicts the experimental results. Applying $U_d=3.5-7.5$ eV, we find the $J_{1r}$ values in the range from $-50$~K to $-10$~K and $J_{2r}=240-360$~K. These numbers are in reasonable agreement with the TB estimates but in conflict with the experimental energy scale. Experimental values of the Curie-Weiss temperature ($\theta=9.6$~K)\cite{kageyama2005} and the saturation field ($\mu_0H_s=30.1$~T)\cite{kageyama2005-2} are quite small. Assuming the frustrated square lattice model with $J_{1r}=-30$~K and $J_{2r}=300$~K, we find $\theta=J_{1r}+J_{2r}=270$~K\cite{rosner2003} and $\mu_0H_s=(2J_{1r}+4J_{2r})k_B/(g\mu_B)\simeq 840$~T,\cite{schmidt2007} in obvious contradiction to the experimental results. Thus, correlations alone are unable to account for the orbital ordering and the resulting spin physics of (CuCl)LaNb$_2$O$_7$. Below, we demonstrate that one has to consider lattice degrees of freedom as well.

\section{Realistic structure}
\label{distorted}
\subsection{Structural distortion}
To derive the structural distortion in (CuCl)LaNb$_2$O$_7$, we use the supercell with doubled $a$ and $b$ parameters, as suggested by the electron diffraction studies.\cite{yoshida2007} We focus on the forces in the $ab$ plane and neglect the forces along the $c$ axis. In Sec.~\ref{methods}, we have noticed the sizable forces along the $c$ axis for Nb and O atoms due to the uncertainty of the experimental structure determination. Such forces are present in both the regular and the distorted structures. Our calculations are performed for two different sets of structural data showing forces of opposite signs. The two sets are opposite with respect to the fully relaxed one and yield essentially similar results. Hence, the precise positions of the atoms along the $c$ axis bear little influence on the structure in the $ab$ plane and on the resulting magnetic interactions.

\begin{figure*}
\includegraphics{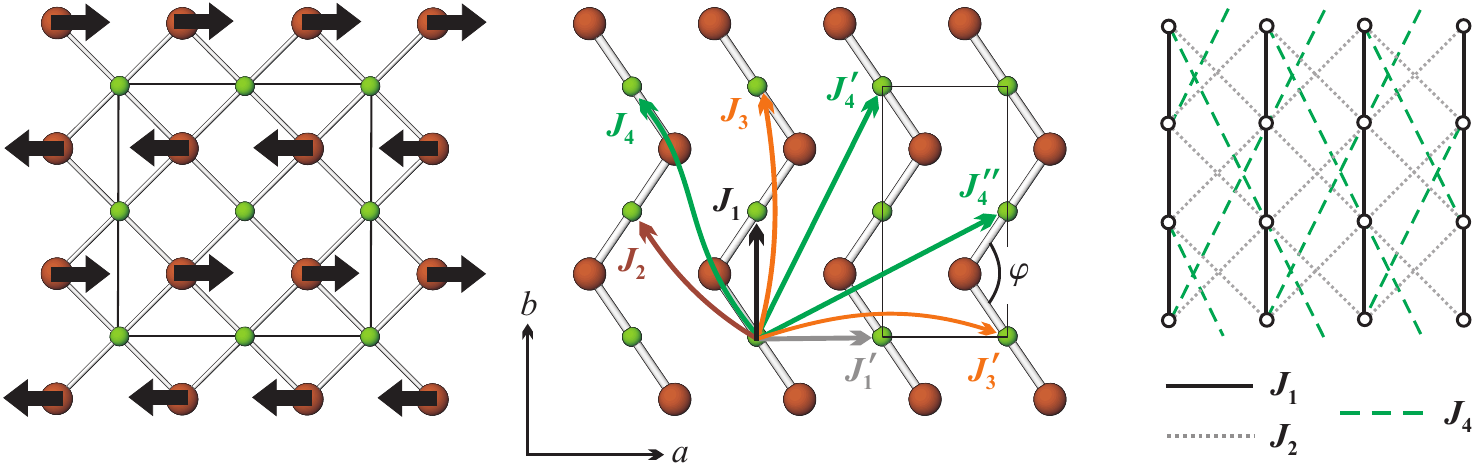}
\caption{\label{fig_distortion}\label{fig_interactions}(Color online)
Structural distortion in the [CuCl] layers: the forces on the Cl atoms (left panel), magnetic interactions in the distorted layer (middle panel), and the resulting spin model (right panel). Smaller and larger spheres show the Cu and Cl atoms, respectively; $\varphi$ measures the Cu--Cl--Cu angle. In the left panel, the square marks the supercell employed in the structure optimization, while in the middle panel the rectangle shows the crystallographic unit cell of the distorted structure. Thick lines in the middle panel are the short Cu--Cl bonds, indicating the CuO$_2$Cl$_2$ plaquettes arranged perpendicular to the layer. In the right panel, only the leading interactions are shown. The solid, dotted, and dashed lines denote $J_1,J_2$, and $J_4$, respectively.
}
\end{figure*}

\begin{figure}
\includegraphics[scale=0.95]{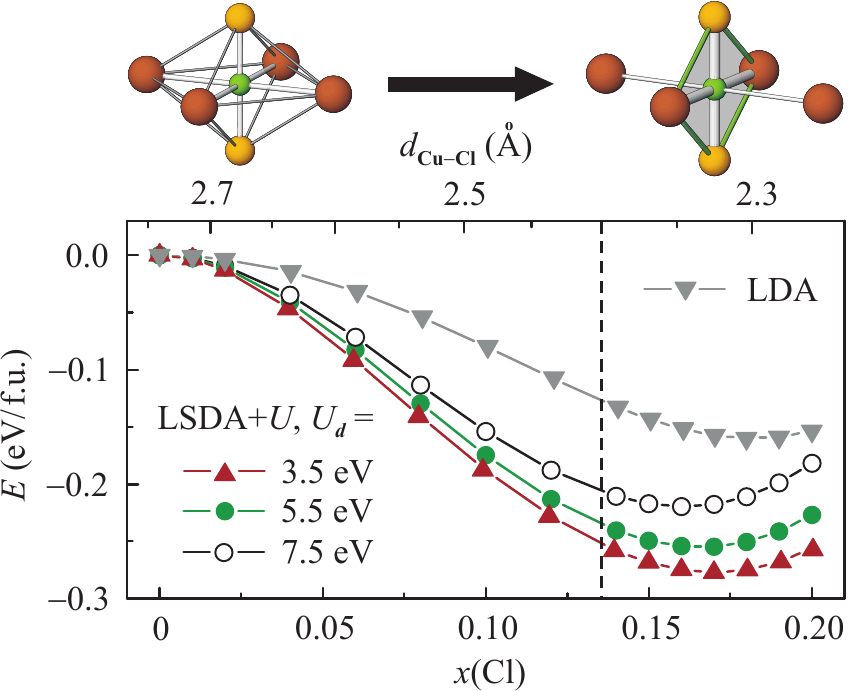}
\caption{\label{fig_energy}(Color online)
The transformation of the copper local environment (upper part) and the change in the total energy (bottom part) along the structural distortion in (CuCl)LaNb$_2$O$_7$. The total energy of the regular (tetragonal) structure is set to zero. The upper axis shows the evolution of the shorter Cu--Cl distance. In the bottom axis, $x$ indicates the coordinate of the $(x,0,\frac12)$ position of the Cl atoms. The $x$ value is measured in units of $a$ (the lattice parameter of the regular structure). The dashed line shows the $x$ coordinate of the experimental four-fold position of Cl with random occupation, as determined by the refinement of the neutron data.\cite{caruntu2002} In the upper panel, the shaded region marks the CuO$_2$Cl$_2$ plaquette.
}
\end{figure}

The regular structure does not reveal any sizable forces in the $ab$ plane. However, the displacement of the Cl atoms away from the high-symmetry position leads to the forces that tend to enhance the distortion. The supercell includes four independent Cl atoms, but the forces on these atoms are found to be similar. This indicates the cooperative character of the distortion. The shifts of the Cl atoms give rise to two short and two long Cu--Cl bonds. The resulting distortion pattern is shown in Fig.~\ref{fig_distortion}. The Cl atoms occupy the $(x,0,\frac12)$ sites in an ordered manner to yield opposite directions for the short Cu--Cl bonds on each Cu site. The structural transformation is quantified by a single parameter, the $x$ value in $(x,0,\frac12)$ (in the following, we assume $x$ to be the coordinate of the shifted Cl atoms in the initial, tetragonal unit cell). The change in the total energy in the course of the distortion is visualized in Fig.~\ref{fig_energy}.

Both LDA and LSDA+$U$ predict the stabilization of the distorted structure with respect to the regular one. The relaxed positions of the Cl atoms are nearly the same in all the calculations, while the stabilization energy ($\Delta E$) shows a pronounced dependence on the computational method and on $U_d$. The $\Delta E$ value is the energy gain due to the formation of two short Cu--Cl bonds in the distorted structure rather than four longer bonds in the regular structure (see the upper part of Fig.~\ref{fig_energy}). Therefore, $\Delta E$ depends on the hybridization between Cu and Cl orbitals, the latter being controlled by the $U_d$ value. The application of the $U_d$ potential shifts filled copper states to lower energy, hence providing stronger hybridization and increasing $\Delta E$ at low $U_d$ (3.5~eV). Higher values of $U_d$ further shift the copper states to lower energy, reduce the hybridization, and $\Delta E$ is decreased. Yet, even the smallest (LDA) value ($\Delta E=0.15$~eV/f.u.) is sufficient to stabilize the distortion with respect to thermal fluctuations (at room temperature, thermal energy amounts to about 0.025~eV). This finding contrasts to the studies of KCuF$_3$ (Ref.~\onlinecite{kcuf3}) and Cs$_2$AgF$_4$ (Ref.~\onlinecite{cs2agf4}), where the distortion is stabilized within LSDA+$U$ only. In (CuCl)LaNb$_2$O$_7$, both the lattice and the correlations favor the distorted structure.

Our relaxed structure is different from one suggested by Yoshida \textit{et~al}.\cite{yoshida2007} Basically, an ordering similar to ours was considered by Whangbo and Dai,\cite{whangbo2006} though they did not optimize the positions of the Cl atoms, did not find the opportunity for the spin-gap behavior and, finally, discarded the model. Neither of the two groups supported the choice of the model by total energy calculations and proved the stability of the distorted structure with respect to the regular one. In contrast to the previous studies, our proposition for the structural model is well-justified and confirmed by the sizable stabilization energies on different levels of theory (Fig.~\ref{fig_energy}). Below, we will demonstrate that our model is also consistent with the available experimental data.

First of all, the shift of the Cl atoms to the $(x,0,\frac12)$ position is in agreement with the neutron diffraction results.\cite{caruntu2002} Despite the lack of the superstructure reflections in the neutron data, it was possible to refine the $x$ value by assuming random occupation of four equivalent $(x,0,\frac12)$ sites. The resulting $x\simeq 0.136$ (dashed vertical line in Fig.~\ref{fig_energy}) is in reasonable agreement with the positions of the energy minima ($x=0.16-0.18$). The optimized structure shows single sites for the Cu, Cl, and La atoms in agreement with the NMR and NQR data.\cite{yoshida2007} Yet, the $a\times 2a\times c$ unit cell (see Fig.~\ref{fig_distortion}) seems to contradict the electron diffraction results, suggesting the doubling of both $a$ and $b$ parameters.\cite{yoshida2007} However, one should keep in mind that in the regular structure the $a$ and $b$ axes are indistinguishable, hence the Cl atoms can be equally shifted to either $(x,0,\frac12)$ or $(0,x,\frac12)$. Then, one expects the $90^{\circ}$ twinning with twins leading to two different sets of superstructure reflections in the electron diffraction patterns and to the apparent doubling along both the $a$ and $b$ axes. Thus, the twinning can explain the seeming controversy between the structural model and the electron diffraction data, although a careful high-resolution electron microscopy (HREM) study is desirable to resolve this issue.

We find a further experimental confirmation for the proposed structural model by considering EFG's. The EFG's for the Cu, Cl, and La sites were measured experimentally by NMR on aligned powder samples.\cite{yoshida2007} Therefore, several characteristics can be determined: (i) the leading principal component of the tensor ($V_{zz}$); (ii) the asymmetry parameter [$\eta=(V_{xx}-V_{yy})/V_{zz}$]; and (iii) the orientation of the $z$ axis of the tensor with respect to the $c$ axis of the crystal structure. The EFG's are a sensitive tool to detect the local symmetry of an atomic position. In tetragonal structures, the positions on the four-fold axes should have $V_{xx}=V_{yy}$, hence $\eta=0$. The non-zero experimental $\eta$ value for the Cu, Cl, and La sites (see the last line of Table~\ref{tab_efg}) is a strong evidence for the structural distortion in (CuCl)LaNb$_2$O$_7$.\cite{yoshida2007} 

The calculated EFG parameters for the regular and distorted structures are listed in Table~\ref{tab_efg} along with the experimental results. The regular structure suggests $\eta=0$ for all the three sites and $z\!\parallel\! c$ for the Cu site in clear contradiction to the experiment. Yet, the distorted structure provides reasonable $\eta$ values for the Cu and Cl sites. For the Cu site, $V_{zz}$ and $\eta$ strongly depend on $U_d$, since the on-site correlations modify the electronic distribution around this site. For the Cl atoms, the dependence on $U_d$ is weak, while the results for the La site are almost independent on $U_d$, because the La and Cu sites are far apart. At $U_d=5.5$~eV and 7.5~eV, the EFG's on the Cu and Cl sites are in good agreement with the experiment. The orientation of the EFG tensors is also reproduced. For the La site, the calculations yield the reasonable value of $V_{zz}$ and the correct orientation of the tensor. However, the calculated asymmetry parameter is well below the experimental one.

\begin{table}
\caption{\label{tab_efg}
Computational estimates of the electric field gradients for the distorted and regular structures of (CuCl)LaNb$_2$O$_7$: leading principal components of the tensors $V_{zz}$ (in $10^{21}$ V/m$^2$), asymmetries $\eta$, and orientations of the tensor ($z$ vs. $c$). $U_d$ is the Coulomb repulsion parameter of LSDA+$U$. The bottom part of the table lists the experimental results from Ref.~\onlinecite{yoshida2007}.
}
\begin{ruledtabular}
\begin{tabular}{c@{\hspace{4ex}}cc@{\hspace{4ex}}cc@{\hspace{4ex}}cc}
  & \multicolumn{2}{c}{Cu} & \multicolumn{2}{c}{Cl} & \multicolumn{2}{c}{La} \\
  $U_d$ (eV) & $V_{zz}$ & $\eta$ & $V_{zz}$ & $\eta$ & $V_{zz}$ & $\eta$ \\\hline
  Distorted structure & \multicolumn{2}{c}{$z\perp c$} & \multicolumn{2}{c}{$z\!\parallel\! c$} & \multicolumn{2}{c}{$z\!\parallel\! c$} \\
  3.5 & $-6.9$ & 0.38 & $-19.1$ & 0.62 & 5.6 & 0.03 \\
  5.5 & $-9.8$ & 0.05 & $-18.0$ & 0.60 & 5.6 & 0.03 \\
  7.5 & $-12.2$ & 0.20 & $-17.1$ & 0.59 & 5.7 & 0.03 \\\hline
  Regular structure & \multicolumn{2}{c}{$z\!\parallel\! c$} & \multicolumn{2}{c}{$z\!\parallel\! c$} & \multicolumn{2}{c}{$z\!\parallel\! c$} \\ 
  3.5 & 9.0 & 0 & $-18.4$ & 0 & 5.5 & 0 \\
  5.5 & 12.7 & 0 & $-17.3$ & 0 & 5.5 & 0 \\
  7.5 & 15.4 & 0 & $-16.5$ & 0 & 5.6 & 0 \\\hline
  Experiment\footnotemark & \multicolumn{2}{c}{$z\perp c$} & \multicolumn{2}{c}{$z\!\parallel\! c$} & \multicolumn{2}{c}{$z\!\parallel\! c$} \footnotetext{Note that the NMR data do not allow to determine the sign of $V_{zz}$. The values are listed according to Ref.~\onlinecite{yoshida2007}, though the assignment is basically arbitrary.}\\
  & 11.6 & 0.10 & $-14.2$ & 0.56 & 6.5 & 0.70 \\
\end{tabular}
\end{ruledtabular}
\end{table}

\begin{figure}
\includegraphics{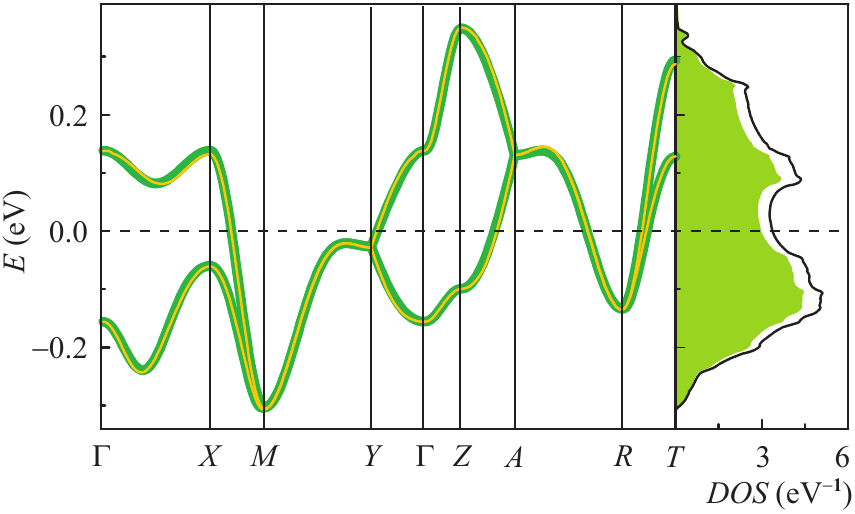}
\caption{\label{fig_bandd}(Color online)
LDA bands for the distorted structure of (CuCl)LaNb$_2$O$_7$ (left part) and the respective density of states for copper atoms (right part). The Fermi level is at zero energy. The Cl atoms occupy the $(x,0,\frac12)$ position with $x=0.16$. In the left panel, the thin (light) lines denote the LDA bands, while the thick (dark) lines show the fit of the tight-binding model. In the right panel, the solid line indicates the total density of states for copper atoms, and the solid filling marks the contribution of the in-plaquette orbitals. The notation of $k$ points is as follows: $\Gamma(0,0,0)$, $X(0.5,0,0)$, $M(0.5,0.5,0)$, $Y(0,0.5,0)$, $Z(0,0,0.5)$, $A(0,0.5,0.5)$, $R(0.5,0.5,0.5)$, and $T(0.5,0,0.5)$ (the coordinates are given along the $k_x,k_y$, and $k_z$ axes in units of the respective reciprocal lattice parameters $2\pi/a$, $\pi/a$, and $2\pi/c$).
}
\end{figure}
To understand the discrepancy for the asymmetry of the La site EFG, one should recall that the low asymmetry implies the weak distortion of the local environment with respect to the regular structure (for the regular structure $\eta=0$, see Table~\ref{tab_efg}). Our structural model suggests the distortion of the [CuCl] layers, while the [LaNb$_2$O$_7$] blocks retain the tetragonal symmetry. Therefore, the low calculated $\eta$ value is reasonable. The experimental $\eta=0.70$ implies the distortion of the [LaNb$_2$O$_7$] blocks. Such a distortion may take place due to a tilting of the NbO$_6$ octahedra. The tilting distortions are typical for perovskite-type structures,\cite{wells,woodward} and one can expect this type of the distortion in (CuCl)LaNb$_2$O$_7$.\cite{foot5} However, the band structure calculations do not evidence any changes in the [LaNb$_2$O$_7$] blocks, since the $ab$ components of the forces on the Nb and O atoms are negligible (below 0.01 eV/a.u.) The high-symmetry structure may correspond to a local energy maximum, where the forces should vanish. However, the shifts of the oxygen atoms away from the high-symmetry positions yield sizable forces that tend to restore the high-symmetry structure. Thus, there is no evidence for the distortion of the [LaNb$_2$O$_7$] blocks. Possibly, one has to use a larger supercell in order to observe this distortion. However, the electron diffraction data\cite{yoshida2007} do not suggest a larger supercell leaving the distortion of the [LaNb$_2$O$_7$] blocks as an open question. 

Finally, one finds support for the distorted structure by considering the crystal chemistry of copper compounds. The distortion leads to the planar CuO$_2$Cl$_2$ environment of copper (see the upper part of Fig.~\ref{fig_energy}) with two Cu--Cl bonds of $2.3-2.4$ \r A. Two other \mbox{Cu--Cl} distances are extended up to 3.2 \r A and become non-bonding. The CuO$_2$Cl$_2$ planar coordination is typical for copper oxychlorides, and the Cl atoms are normally located in the opposite corners of the O$_2$Cl$_2$ rectangle similar to our model (Fig.~\ref{fig_distortion}).\cite{wells} Therefore, the distortion in (CuCl)LaNb$_2$O$_7$ is also in accord with the empirical structural expectations.

Basically, the CuO$_2$Cl$_2$ unit can be considered as an analog of the conventional CuO$_4$ plaquette. The CuO$_2$Cl$_2$ plaquette is geometrically distorted due to the different Cu--O and Cu--Cl bond lengths. However, its electronic structure is similar to that of the CuO$_4$ square. The half-filled orbital lies in the CuO$_2$Cl$_2$ plane, and the lobes of this orbital point to both the O and Cl atoms, as illustrated by the respective Wannier function (see the left panel of Fig.~\ref{fig_wannier}). The Wannier function also shows sizable contributions from $\sigma$-type O and Cl orbitals. The position of the copper orbital is in agreement with intuitive crystal-field considerations. LDA reveals two bands at the Fermi level, and these bands are predominantly formed by the in-plaquette orbitals of two copper atoms (see Fig.~\ref{fig_bandd}). Thus, the structural distortion lifts the orbital degeneracy and leads to the single-orbital ground state typical for most of the Cu$^{+2}$ compounds.

\subsection{Exchange couplings}
In this section, we evaluate the leading exchange couplings for the distorted crystal structure of (CuCl)LaNb$_2$O$_7$ and derive a realistic spin model for this compound. Below, we use two different $(x,0,\frac12)$ positions of the Cl atoms: $x=0.16$ as a representative position for the optimized structure (see Fig.~\ref{fig_energy}) and $x=0.136$ as the coordinate suggested by the neutron diffraction experiment.\cite{caruntu2002} We also consider general trends for the evolution of the exchange couplings along the shift of the Cl atom. 

\begin{table*}
\caption{\label{tab_hoppings}
Leading hopping parameters ($t_i$) of the tight-binding model and the resulting antiferromagnetic contributions to the exchange couplings ($J_i^{\AFM}=4t_i^2/U_{\eff}$) for two different positions of the Cl atoms: $x=0.136$ (experimental) and $x=0.16$ (optimized). For the on-site Coulomb repulsion potential, $U_{\eff}=4.5$~eV has been used a representative value.
}
\begin{ruledtabular}
\begin{tabular}{cccccccccc}
  $t$ (meV) & $t_1$ & $t_1'$ & $t_2$ & $t_3$ & $t_3'$ & $t_4$ & $t_4'$ & $t_4''$ & $t_{\perp}$ \\
  $x=0.136$ & $51$ & $-22$ & $26$ & $-16$ & $4$ & $42$ & $0$ & $-1$ & $39$ \\
  $x=0.16$ & $70$ & $-23$ & $18$ & $-22$ & $3$ & $44$ & $0$ & $-1$ & $41$ \\\hline
  $J^{\AFM}$ (K) & $J_1$ & $J_1'$ & $J_2$ & $J_3$ & $J_3'$ & $J_4$ & $J_4'$ & $J_4''$ & $J_{\perp}$ \\
  $x=0.136$ & $27$ & $5$ & $7$ & $3$ & $0.2$ & $18$ & $0$ & $<0.1$ & $16$ \\
  $x=0.16$ & $51$ & $5$ & $3$ & $5$ & $0.1$ & $20$ & $0$ & $<0.1$ & $17$ \\
\end{tabular}
\end{ruledtabular}
\end{table*}

The distortion lifts the tetragonal symmetry and leads to a large number of inequivalent superexchange pathways (Fig.~\ref{fig_interactions}). The distorted layer can be considered as a system of corner-sharing CuO$_2$Cl$_2$ plaquettes forming chains along the $b$ axis. Then, the single nearest-neighbor interaction of the regular structure ($J_{1r}$, see Fig.~\ref{fig_structure}) is split into two interactions: $J_1$ runs along the chains and $J_1'$ runs perpendicular to the chains. The next-nearest-neighbor coupling $J_2$ is unique, similar to the regular structure. Third-neighbor interactions are again split into two interactions: $J_3$ runs along the chains and $J_3'$ runs perpendicular to the chains. Finally, there are three inequivalent fourth-neighbor couplings $J_4,J_4'$, and $J_4''$. In the following, we select the relevant interactions by analyzing the LDA band structure. Then, we also evaluate these interactions via the LSDA+$U$ calculations.

\begin{figure}
\includegraphics[scale=0.95]{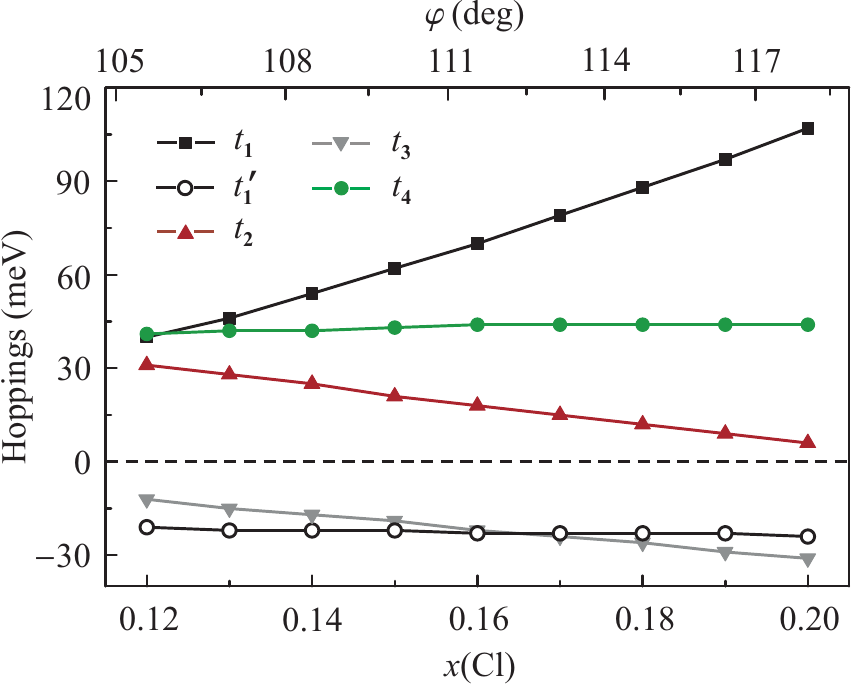}
\caption{\label{fig_trends}
The evolution of the leading hoppings parameters within the [CuCl] layer upon shifting the Cl atom. The $x$ value is measured in units of $a$ (the lattice parameter of the regular structure) and indicates the coordinate of the $(x,0,\frac12)$ position of the Cl atoms. The upper axis shows the change in the Cu--Cl--Cu angle relevant for the interaction $J_1$ ($\varphi$, see the middle panel of Fig.~\ref{fig_distortion}).
}
\end{figure}

The LDA band structure is shown in Fig.~\ref{fig_bandd}. Two bands at the Fermi level correspond to two Cu atoms per doubled unit cell in contrast to Fig.~\ref{fig_bandt} with two bands, originating from two orbitals of the same Cu atom in the original unit cell. The bands in Fig.~\ref{fig_bandd} are fitted with a TB model. The parameters of this model (hoppings) are calculated as overlap integrals of the Wannier functions centered on copper sites. The resulting numbers are listed in Table~\ref{tab_hoppings}. The largest hopping runs along the chains of the CuO$_2$Cl$_2$ plaquettes and leads to the AFM interaction $J_1$ of about 30~K or 50~K depending on the position of the Cl atoms. Other in-layer hoppings are relatively small, and the second largest hopping is the interlayer term ($t_{\perp}$) that yields the AFM interlayer interaction of about 15~K. There is a pronounced difference between the inequivalent fourth-neighbor interactions. One of these interactions is comparable to $J_{\perp}$ ($J_4^{\AFM}\simeq 20$~K), while others are negligible. The interactions beyond the fourth neighbors are also negligible (the respective hoppings are below 10~meV implying $J_i^{\AFM}$ below 1~K). Thus, the TB fit suggests the following model for the spin system of (CuCl)LaNb$_2$O$_7$: AFM chains run along the $b$ direction, and these chains are coupled by the AFM interactions $J_4$ and $J_{\perp}$ (see the right panel of Fig.~\ref{fig_distortion}). The evolution of the leading hoppings parameters along the shift of the Cl atom is shown in Fig.~\ref{fig_trends}.

The TB estimates are in qualitative agreement with the experimental data. The largest AFM coupling does not exceed 50 K and shows the same order of magnitude as the Curie-Weiss temperature\cite{kageyama2005} and the saturation field (see Sec.~\ref{tetragonal}).\cite{kageyama2005-2} This strongly supports the structural distortion and the proposed spin model in contrast to the frustrated square lattice scenario suggested by the regular structure. The sizable long-range interaction $J_4$ is in agreement with the INS data.\cite{kageyama2005} In Sec.~\ref{discussion}, we will further compare our spin model with the experimental findings. Now, we proceed to the LSDA+$U$ results that yield supplementary information on the leading exchange couplings in (CuCl)LaNb$_2$O$_7$.

\begin{figure}
\includegraphics[scale=0.9]{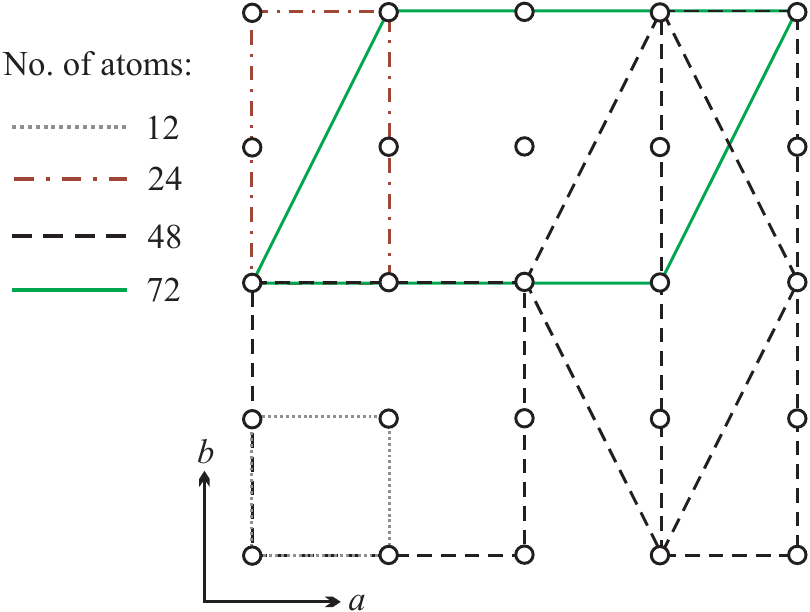}
\caption{\label{fig_scells}(Color online)
Different unit cells and supercells employed in the present work. Small spheres show the Cu atoms. The dotted line indicates the 12-atom unit cell of the regular (tetragonal) structure of (CuCl)LaNb$_2$O$_7$. The dash-dotted line shows the 24-atom unit cell of the distorted structure. The dashed and solid lines denote the 48- and 72-atom supercells employed in the LSDA+$U$ calculations.
}
\end{figure}

For the LSDA+$U$ calculations, we select all the in-layer interactions exceeding 1~K, i.e., $J_1,J_1',J_2,J_3$, and $J_4(J_4')$. To evaluate these couplings, one has to use a large, 96-atom supercell that makes full-potential calculations quite time-consuming and, likely, not very accurate. Therefore, we employ an alternative approach and consider smaller cells. Three 48-atom supercells (Fig.~\ref{fig_scells}) enable to evaluate a number of linear combinations of $J_1,J_1',J_2,J_3$, and $(J_4+J_4')$. Then the individual couplings are evaluated by solving a simple system of linear equations. We also use one 72-atom supercell (Fig.~\ref{fig_scells}) to distinguish between $J_4$ and $J_4'$. The calculations for four different supercells are basically redundant, and some quantities are independently calculated within two different supercells. The respective estimates match within 2~K and indicate the internal consistency and completeness of our LSDA+$U$ results.

\begin{table}
\caption{\label{tab_lsda+u}
LSDA+$U$ estimates of the exchange couplings $J_i$ (in~K) in the distorted structure of (CuCl)LaNb$_2$O$_7$ (the positions of the Cl atoms are fixed at $x=0.16$). $U_d$ (in eV) is the Coulomb repulsion parameter of LSDA+$U$.
}
\begin{ruledtabular}
\begin{tabular}{ccccccc}
  $U_d$ & $J_1$ & $J_1'$ & $J_2$ & $J_3$ & $J_4$ & $J_4'$ \\
  $3.5$ & $51$ & $-2$ & $-5$ & $9$ & $39$ & $7$ \\
  $5.5$ & $63$ & $0$ & $18$ & $1$ & 29 & $2$ \\
  $7.5$ & $64$ & $-2$ & $29$ & $2$ & $20$ & $0$ \\
\end{tabular}
\end{ruledtabular}
\end{table}

The LSDA+$U$ estimates of individual exchange couplings are listed in Table~\ref{tab_lsda+u}. The calculations were done for $x=0.16$ only. The results are in good agreement with the TB estimates. The leading AFM interaction is $J_1$, whereas the interactions $J_1'$ and $J_3$ are weak consistent with the small hoppings $t_1'$ and $t_3$, and $J_4$ is drastically different from $J_4'$. Ferromagnetic contributions to the exchange couplings are small. The values of $J_2$ and $J_4$ depend on the specific choice of the $U_d$ parameter: at $U_d=7.5$~eV, $J_2$ is heavily overestimated with respect to $J_2^{\AFM}$ (Table~\ref{tab_hoppings}). On the other hand, the low $U_d$ of 3.5~eV leads to a somewhat overestimated $J_4$. Thus, the intermediate $U_d$ value of 5.5~eV seems to be the optimal one, suggesting similar AFM interactions $J_2$ and $J_4$ that amount to approximately half of $J_1$. Basically, the choice of $U_d=5.5$~eV also looks reasonable from the microscopic point of view. The $U_d$ values of $6.5-8.0$~eV are normally used for the LSDA+$U$ calculations of Cu$^{+2}$-containing oxides with copper surrounded by oxygen atoms only, while Cl $3p$ orbitals provide better screening of the on-site repulsion thus reducing $U_d$.\cite{foot4}

The resulting exchange couplings can be rationalized as follows. In the distorted structure, the basic structural element is the CuO$_2$Cl$_2$ plaquette. The strongest interaction $J_1$ runs between the corner-sharing plaquettes via the Cu--Cl--Cu superexchange pathway. The magnitude of this interaction depends on the respective hopping, while the hopping is controlled by the Cu--Cl--Cu angle ($\varphi$, see Fig.~\ref{fig_distortion}) and the position of the Cl atom. In the regular structure, $\varphi=90^{\circ}$, the hopping $t_{1r}$ is close to zero, and the nearest-neighbor interaction is weak FM (see Sec.~\ref{tetragonal}). The shift of the Cl atoms leads to the increase in the $\varphi$ angle and to the increase in the hopping $t_1$ (see Fig.~\ref{fig_trends}). Then, the sign of $J_1$ is changed and the interaction becomes AFM. At sufficiently large distortion, $J_1$ is the leading AFM interaction. Yet, $J_1$ in the distorted structure is one order of magnitude below $J_{2r}$ in the regular structure, since the $\varphi$ angle for $J_1$ is still close to $90^{\circ}$ (e.g., $\varphi=111.5^{\circ}$ at $x=0.16$), while $J_{2r}$ corresponds to the linear Cu--Cl--Cu superexchange pathway ($\varphi=180^{\circ}$). 

\begin{figure}
\includegraphics{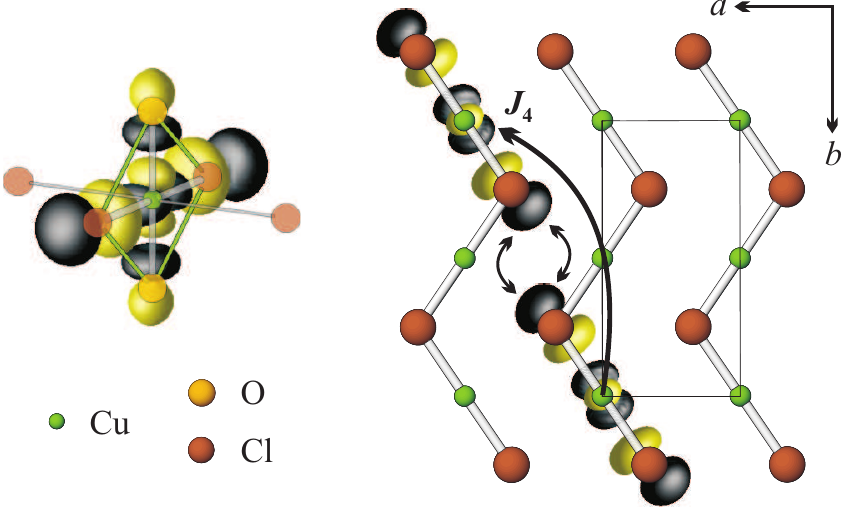}
\caption{\label{fig_wannier}
Plots of the Wanner functions for the distorted structure of (CuCl)LaNb$_2$O$_7$ with $x=0.16$. Left panel: the single CuO$_2$Cl$_2$ plaquette. Right panel: the Cu--Cl--Cl--Cu superexchange pathway causing the large fourth neighbor interaction~$J_4$. 
}
\end{figure}

In contrast to $t_1$, the shift of the Cl atoms does not change $t_1'$, because the respective Cu--Cl--Cu superexchange pathway includes one long, non-bonding Cu--Cl distance, implying the very weak overlap of the Cu and Cl orbitals. The small hopping $t_1'$ suggests a possible FM interaction $J_1'$. The next-nearest-neighbor hopping is also small due to the lack of the proper superexchange pathway. This interaction is reduced with the increase in the $\varphi$ angle. Yet, some long pathways (about 9~\r A and 12~\r A for $J_4$ and $J_{\perp}$, respectively) are quite efficient. The origin of these long-range couplings can be understood by considering the Wannier functions centered on copper sites (see Fig.~\ref{fig_wannier}). The single Wannier function is formed by the in-plaquette $d$ orbital of copper with large contributions from $\sigma$-type orbitals of oxygen and chlorine ($2p$ and $3p$, respectively). The oxygen $p$ orbitals further overlap with niobium $4d$ orbitals, hence the interlayer coupling $J_{\perp}$ should involve the complex Cu--O--Nb--O--Nb--O--Cu superexchange pathway. In case of $J_4$, the interaction is likely mediated by the direct Cl--Cl contact, as indicated by the overlap of the Wannier functions in the right panel of Fig.~\ref{fig_wannier}. This regime can be understood as the $\sigma$-overlap of the Wannier functions, because their lobes point to each other. The similar coupling regime has been reported for the Cu$_2$Te$_2$O$_5$X$_2$ (X = Cl, Br)\cite{cute2252} compounds that show sizable long-range interactions mediated by the X--X contacts (the respective hoppings are about 80~meV). In contrast, the pathways $J_4'$ and $J_4''$ do not lead to the effective overlap of the Wannier function and correspond to the very weak exchange. The same holds for another tellurium-containing copper oxychloride Cu$_4$Te$_5$O$_{12}$Cl$_4$ (Ref.~\onlinecite{cute45124}) where the Cl--Cl contact leads to the small hopping $t_d<10$ meV.

It is worth to note that the long-range interaction $J_4$ is basically independent on the precise Cl position within the studied range of $x=0.12-0.20$ (see Fig.~\ref{fig_trends}). On the other hand, the regular structure does not lead to any sizable fourth neighbor interactions. Thus, there should be a crossover regime at weak distortions. In fact, the scenario of the small distortions is even more complex, because the reduction in $x$ leads to the enhanced fifth neighbor interaction (at $x=0.12$, $t_5$ is about 30~meV). This scenario may be relevant for Br-containing compounds that show smaller distortions as compared to the Cl-containing counterparts. The respective discussion lies beyond the scope of the present work and should be a subject of a further study.

\section{Discussion}
\label{discussion}
The spin gap behavior and the lack of the long-range magnetic ordering in (CuCl)LaNb$_2$O$_7$ have remained a puzzle since the first report of Kageyama \textit{et~al}.\cite{kageyama2005} appeared. Further experimental studies\cite{yoshida2007} indicated a structural distortion, but the distorted structure remained unknown. Two models were suggested;\cite{whangbo2006,yoshida2007} however, neither of them was supported by total energy calculations and quantitative analysis of the exchange couplings. Our study presents a different structural model which is proposed and justified by careful full-potential band structure calculations. We argue that this model is able to explain the unusual magnetic behavior of (CuCl)LaNb$_2$O$_7$.

Structural distortion is a key to understanding the properties of (CuCl)LaNb$_2$O$_7$. The distortion is realized via the cooperative shifts of the Cl atoms. The shifts lead to the formation of short Cu--Cl bonds and the resulting CuO$_2$Cl$_2$ plaquettes. The plaquettes are arranged perpendicular to the [CuCl] layers and share corners to form chains (Fig.~\ref{fig_distortion}). The origin of the distortion deals with the tendency to lift the orbital degeneracy of copper. The regular, tetragonal structure of (CuCl)LaNb$_2$O$_7$ leads to two competing copper orbitals near the Fermi level, while the distortion stabilizes the single-orbital ground state with the half-filled in-plaquette orbital (Fig.~\ref{fig_bandd}). Then, the resulting physics is similar to that of conventional cuprates with CuO$_4$ square-planar structural units. The overall behavior resembles copper compounds with the cooperative Jahn-Teller distortion (e.g., KCuF$_3$). However, the magnitude of the distortion is different, and the driving force is likely different as well. In KCuF$_3$, the shifts of the ligands are small (below 0.2 \r A), while in (CuCl)LaNb$_2$O$_7$ the Cl atoms are shifted for about 0.6~\r A. To stabilize the distorted structure of KCuF$_3$, one has to include the correlations,\cite{kcuf3,kcuf3-2} while the distorted structure of (CuCl)LaNb$_2$O$_7$ is stable even in LDA (Fig.~\ref{fig_energy}). 

The structural distortion has a pronounced effect on the magnetic interactions. The regular structure corresponds to the two-dimensional, frustrated square lattice model, while the distortion leads to a more complex spin model with numerous inequivalent exchange couplings. This model can be understood as one-dimensional in case only the leading coupling $J_1$ is considered. Yet, the introduction of the interchain couplings leads to a three-dimensional spin model, since $J_2$, $J_4$, and $J_{\perp}$ are of similar magnitude. Band structure calculations yield numerical estimates for individual exchange couplings, and it is instructive to compare these estimates with the experimental data.

In the previous section, we have emphasized that the calculated $J$ values have the same order of magnitude as the relevant experimental numbers, the Curie-Weiss temperature ($\theta$) and the saturation field ($\mu_0H_s$). To make a numerical comparison, we employ the spin model shown in the right panel of Fig.~\ref{fig_interactions}. The Curie-Weiss temperature is the coefficient at the second-order term of the high-temperature expansion for the magnetic susceptibility ($\chi$ in powers of $1/T$). Thus, $\theta$ can be expressed as $\theta=1/4\sum_iz_iJ_i$, where $z_i$ is the coordination number for the interaction $J_i$. The $\theta$ value of 9.6~K\cite{kageyama2005} is below the position of the susceptibility maximum ($T_{\max}^{\chi}\simeq 16$~K) and the spin gap ($\Delta\simeq 26$~K).\cite{kageyama2005} Then the low Curie-Weiss temperature is likely caused by the presence of both FM and AFM interactions in (CuCl)LaNb$_2$O$_7$. Our calculations do not show any sizable FM interactions and yield $\theta$ of $40-50$ K, well above the experimental value of 9.6~K.\cite{kageyama2005} Thus, the FM interactions are underestimated. Yet the AFM interactions seem to be overestimated, because the leading AFM interaction $J_1\simeq 50$~K is above the saturation field of 30~T (i.e., 40~K).\cite{kageyama2005-2} However, the $J$ values strongly depend on the position of the Cl atoms, and one may expect a better agreement in case the experimental position is considered. Indeed, at $x=0.136$ the leading coupling $J_1$ is reduced to 30~K (see Table~\ref{tab_hoppings}), i.e., below the value of the saturation field. 

Basically, one has to find out the precise position of the Cl atoms in order to give accurate estimates of the exchange couplings in (CuCl)LaNb$_2$O$_7$. Nevertheless, our spin model derived for the optimized position ($x=0.16$) is qualitatively valid and suggests the plausible mechanism of the magnetic frustration in this material. The right panel of Fig.~\ref{fig_distortion} presents the spin lattice with the leading AFM interactions $J_1,J_2$, and $J_4$. The respective bonds form triangles; therefore, the three interactions can not be satisfied simultaneously. According to the second line of Table~\ref{tab_lsda+u}, $J_2$ and $J_4$ are of similar magnitude, hence a strong frustration of the interchain couplings is expected. Note that the FM interaction $J_1'$ also leads to a frustration via the four-site triangles formed by one $J_1'$, one $J_4$, and two $J_1$ bonds (i.e., one FM and three AFM bonds). 

The strong frustration naturally explains the lack of the long-range ordering in (CuCl)LaNb$_2$O$_7$ despite sizable couplings along both $a,b$, and $c$ axes being present. The explanation of the spin gap behavior is more complicated, because the spin model presented in Fig.~\ref{fig_distortion} has not been studied theoretically. Whangbo and Dai\cite{whangbo2006} reported the classical treatment of a simplified model with $J_1$ and $J_4$ only. They did not find the spin gap, and this result is not surprising, since the $J_1-J_4$ model is non-frustrated. Normally, two-dimensional spin systems without the dimerization are gapless. However, the magnetic frustration can strongly modify the physics and lead to a sizable spin gap as found in, e.g., CaV$_4$O$_9$.\cite{pickett,korotin} We believe that a detailed theoretical treatment of the realistic spin model with at least three interactions ($J_1,J_2$, and $J_4$) is necessary to elucidate the origin of the spin gap in (CuCl)LaNb$_2$O$_7$. Meanwhile, we should note the "unconventional" origin of the spin gap. The gap is not caused by dimerization; therefore, the gap does not simply separate singlet and triplet states of a dimer. Rather, it separates a singlet ground state and a complex excited state of a two-dimensional frustrated spin system. The latter result may be relevant for the explanation of the drastic difference between the spin gap estimates from the high-field magnetization data\cite{kageyama2005-2,kitada2007} and the magnetic susceptibility or INS data.\cite{kageyama2005}

Thus, our calculations suggest a valid spin model that accounts for most of the experimental observations reported so far. The spin model is based on a structural model which is consistent with the available diffraction data and with most of the NMR results. According to Sec.~\ref{distorted}, there is an additional distortion in the [LaNb$_2$O$_7$] block, which for now remains unresolved computationally. However, there is another structural aspect that may also be important. The calculations propose the energetically favorable structural model shown in Fig.~\ref{fig_distortion}. This structure is thermodynamically stable as well, in case the entropy contribution is negligible (this is usually the case in ordered solids at sufficiently low temperatures, e.g., at room temperature). Nevertheless, the compound itself is metastable and readily decomposes at 450~$^{\circ}$C. The preparation temperature is quite low (325~$^{\circ}$C); therefore, the real system does not necessarily reach the equilibrium state proposed by the calculations. Although we expect that the real structure resembles the thermodynamically stable one, the difference between the calculated (equilibrium) and real (non-equilibrium) structures along with the uncertainty of the Cl atom position are the most likely reasons for the discrepancies between the experimental and computational results. 

Unfortunately, there is no way to optimize the structure of non-equilibrium systems. Therefore, our spin model and our estimates for the exchange couplings should be considered as approximations to the properties of the real material. To get further insight, we suggest additional experimental studies. First of all, careful electron microscopy investigations are able to provide supplementary information on the structure of (CuCl)LaNb$_2$O$_7$. These studies should challenge the proposed structural model and investigate the local structure, since a number of defects (e.g., twinning) are expected. It is also desirable to use the experimental data to quantify the spin model proposed by our calculations. The Curie-Weiss temperature and the saturation field hold a certain quantitative information on the exchange couplings. Yet, the INS data should be more informative. In Ref.~\onlinecite{kageyama2005}, the INS data are analyzed within a simple dimer model. Further analysis within the realistic spin model (Fig.~\ref{fig_interactions}) will be very helpful. 

Finally, we suggest that the scenario of the structural distortion may be generic for the whole family of layered perovskite-type copper compounds [(CuX)LaM$_2$O$_7$, (CuX)M$_2'$M$_3$O$_{10}$, etc. with X = Cl, Br; M = Nb, Ta; and M$'$ = Ca, Sr]. All these materials present an interesting but presently poorly understood physics. For example, (CuBr)LaNb$_2$O$_7$ reveals columnar AFM ordering at the unexpectedly high temperature of 32~K (despite the exchange couplings are also of the order of $30-40$~K),\cite{oba2006} while some of the (CuBr)M$_2'$M$_3$O$_{10}$ compounds show a magnetization plateau at 1/3 of the saturation.\cite{tsujimoto2007,tsujimoto2008} The common feature of this compound family is the high-symmetry environment of copper and the tendency of the X atom to be shifted away from the high-symmetry position, as evidenced by high Debye-Waller factors in crystal structure refinements.\cite{kodenkandath1999,kodenkandath2001,tsujimoto2008} We propose that all the compounds should in fact show a structural distortion of Jahn-Teller type. Then, shorter Cu--X bonds will lead to peculiar magnetic interactions, including long-range couplings. The first indication for the relevance of this scenario is found in the recent NMR study of (CuBr)LaNb$_2$O$_7$.\cite{yoshida2008} It is shown that the bromine compound lacks the tetragonal symmetry similar to (CuCl)LaNb$_2$O$_7$. This may suggest a distortion pattern similar to that in the chlorine compound (Fig.~\ref{fig_distortion}). However, the different size of bromine will lead to a different position of the X atoms, hence drastically changing the exchange couplings and the resulting physics. Further computational and experimental studies of the problem are desirable.

In conclusion, we have presented a microscopic model for the structural distortion and the resulting magnetic interactions in (CuCl)LaNb$_2$O$_7$. The model is based on ab-initio electronic structure calculations, conforms to the experimental data (x-ray and neutron diffraction, NMR, and NQR), and explains the peculiar magnetic properties of the compound. We argue that the tetragonal crystal structure leads to a competition of copper orbitals, while the structural distortion lifts this competition and yields a single-orbital ground state. The pronounced distortion is stabilized by both lattice and correlation effects. The distortion gives rise to a drastic change in the magnetic interactions in (CuCl)LaNb$_2$O$_7$. The lack of the tetragonal symmetry and the long-range interactions lead to a complex spin model with chains running along one of the directions in the $ab$ plane and sizable interchain couplings. The interchain interactions in the $ab$ plane are frustrated and may account for the spin-gap behavior.

\begin{acknowledgments}
The authors thank Stefan-Ludwig Drechsler and Artem Abakumov for fruitful discussions, Bella Lake for stimulating our interest to the problem, Katrin Koch and Klaus Koepernik for implementing the EFG module and the Wannier functions in FPLO. Financial support of GIF (Grant No. I-811-257.14/03), RFBR (Project No. 07-03-00890), and the Emmy-Noether-Program of the DFG is acknowledged. A. Ts. is grateful to MPI CPfS for hospitality and financial support during the stay.
\end{acknowledgments}


\begin{thebibliography}{48}
\expandafter\ifx\csname natexlab\endcsname\relax\def\natexlab#1{#1}\fi
\expandafter\ifx\csname bibnamefont\endcsname\relax
  \def\bibnamefont#1{#1}\fi
\expandafter\ifx\csname bibfnamefont\endcsname\relax
  \def\bibfnamefont#1{#1}\fi
\expandafter\ifx\csname citenamefont\endcsname\relax
  \def\citenamefont#1{#1}\fi
\expandafter\ifx\csname url\endcsname\relax
  \def\url#1{\texttt{#1}}\fi
\expandafter\ifx\csname urlprefix\endcsname\relax\def\urlprefix{URL }\fi
\providecommand{\bibinfo}[2]{#2}
\providecommand{\eprint}[2][]{\url{#2}}

\bibitem[{\citenamefont{Kugel and Khomskii}(1982)}]{kugel1982}
\bibinfo{author}{\bibfnamefont{K.~I.} \bibnamefont{Kugel'}} \bibnamefont{and}
  \bibinfo{author}{\bibfnamefont{D.~I.} \bibnamefont{Khomskii}},
  \bibinfo{journal}{Sov. Phys. Usp.} \textbf{\bibinfo{volume}{25}},
  \bibinfo{pages}{231} (\bibinfo{year}{1982}).

\bibitem[{\citenamefont{Sage et~al.}(2006)\citenamefont{Sage, Blake,
  Nieuwenhuys, and Palstra}}]{smvo3}
\bibinfo{author}{\bibfnamefont{M.~H.} \bibnamefont{Sage}},
  \bibinfo{author}{\bibfnamefont{G.~R.} \bibnamefont{Blake}},
  \bibinfo{author}{\bibfnamefont{G.~J.} \bibnamefont{Nieuwenhuys}},
  \bibnamefont{and} \bibinfo{author}{\bibfnamefont{T.~T.~M.}
  \bibnamefont{Palstra}}, \bibinfo{journal}{Phys. Rev. Lett.}
  \textbf{\bibinfo{volume}{96}}, \bibinfo{pages}{036401}
  (\bibinfo{year}{2006}); cond-mat/0511744.

\bibitem[{\citenamefont{Miyasaka et~al.}(2007)\citenamefont{Miyasaka, Yasue,
  Fujioka, Yamasaki, Okimoto, Kumai, Arima, and Tokura}}]{dyvo3}
\bibinfo{author}{\bibfnamefont{S.}~\bibnamefont{Miyasaka}},
  \bibinfo{author}{\bibfnamefont{T.}~\bibnamefont{Yasue}},
  \bibinfo{author}{\bibfnamefont{J.}~\bibnamefont{Fujioka}},
  \bibinfo{author}{\bibfnamefont{Y.}~\bibnamefont{Yamasaki}},
  \bibinfo{author}{\bibfnamefont{Y.}~\bibnamefont{Okimoto}},
  \bibinfo{author}{\bibfnamefont{R.}~\bibnamefont{Kumai}},
  \bibinfo{author}{\bibfnamefont{T.}~\bibnamefont{Arima}}, \bibnamefont{and}
  \bibinfo{author}{\bibfnamefont{Y.}~\bibnamefont{Tokura}},
  \bibinfo{journal}{Phys. Rev. Lett.} \textbf{\bibinfo{volume}{99}},
  \bibinfo{pages}{217201} (\bibinfo{year}{2007}).

\bibitem[{\citenamefont{Zhou et~al.}(2007)\citenamefont{Zhou, Conner, Balicas,
  and Wiebe}}]{sr2vo4}
\bibinfo{author}{\bibfnamefont{H.~D.} \bibnamefont{Zhou}},
  \bibinfo{author}{\bibfnamefont{B.~S.} \bibnamefont{Conner}},
  \bibinfo{author}{\bibfnamefont{L.}~\bibnamefont{Balicas}}, \bibnamefont{and}
  \bibinfo{author}{\bibfnamefont{C.~R.} \bibnamefont{Wiebe}},
  \bibinfo{journal}{Phys. Rev. Lett.} \textbf{\bibinfo{volume}{99}},
  \bibinfo{pages}{136403} (\bibinfo{year}{2007}).

\bibitem[{\citenamefont{Belik et~al.}(2007)\citenamefont{Belik, Iikubo,
  Yokosawa, Kodama, Igawa, Shamoto, Azuma, Takano, Kimoto, Matsui
  et~al.}}]{bimno3}
\bibinfo{author}{\bibfnamefont{A.~A.} \bibnamefont{Belik}},
  \bibinfo{author}{\bibfnamefont{S.}~\bibnamefont{Iikubo}},
  \bibinfo{author}{\bibfnamefont{T.}~\bibnamefont{Yokosawa}},
  \bibinfo{author}{\bibfnamefont{K.}~\bibnamefont{Kodama}},
  \bibinfo{author}{\bibfnamefont{N.}~\bibnamefont{Igawa}},
  \bibinfo{author}{\bibfnamefont{S.}~\bibnamefont{Shamoto}},
  \bibinfo{author}{\bibfnamefont{M.}~\bibnamefont{Azuma}},
  \bibinfo{author}{\bibfnamefont{M.}~\bibnamefont{Takano}},
  \bibinfo{author}{\bibfnamefont{K.}~\bibnamefont{Kimoto}},
  \bibinfo{author}{\bibfnamefont{Y.}~\bibnamefont{Matsui}},
  \bibnamefont{and} \bibinfo{author}{\bibfnamefont{E.}~\bibnamefont{Takayama-Muromachi}},
  \bibinfo{journal}{J. Amer. Chem. Soc.}
  \textbf{\bibinfo{volume}{129}}, \bibinfo{pages}{971} (\bibinfo{year}{2007}).
  
\bibitem[{\citenamefont{Wells}(1975)}]{wells}
\bibinfo{author}{\bibfnamefont{A.~F.} \bibnamefont{Wells}},
  \emph{\bibinfo{title}{Structural Inorganic Chemistry, 4th edition}}
  (\bibinfo{publisher}{Oxford University Press, New York}, \bibinfo{year}{1975}).

\bibitem[{\citenamefont{Khomskii and Kugel}(1973)}]{khomskii1973}
\bibinfo{author}{\bibfnamefont{D.~I.} \bibnamefont{Khomskii}} \bibnamefont{and}
  \bibinfo{author}{\bibfnamefont{K.~I.} \bibnamefont{Kugel}},
  \bibinfo{journal}{Solid State Comm.} \textbf{\bibinfo{volume}{13}},
  \bibinfo{pages}{763} (\bibinfo{year}{1973}).

\bibitem[{\citenamefont{Kodenkandath et~al.}(1999)\citenamefont{Kodenkandath,
  Lalena, Zhou, Carpenter, Sangregorio, Falster, Simmons~Jr., O'Connor, and
  Wiley}}]{kodenkandath1999}
\bibinfo{author}{\bibfnamefont{T.~A.} \bibnamefont{Kodenkandath}},
  \bibinfo{author}{\bibfnamefont{J.~N.} \bibnamefont{Lalena}},
  \bibinfo{author}{\bibfnamefont{W.~L.} \bibnamefont{Zhou}},
  \bibinfo{author}{\bibfnamefont{E.~E.} \bibnamefont{Carpenter}},
  \bibinfo{author}{\bibfnamefont{C.}~\bibnamefont{Sangregorio}},
  \bibinfo{author}{\bibfnamefont{A.~U.} \bibnamefont{Falster}},
  \bibinfo{author}{\bibfnamefont{W.~B.} \bibnamefont{Simmons~Jr.}},
  \bibinfo{author}{\bibfnamefont{C.~J.} \bibnamefont{O'Connor}},
  \bibnamefont{and} \bibinfo{author}{\bibfnamefont{J.~B.} \bibnamefont{Wiley}},
  \bibinfo{journal}{J. Amer. Chem. Soc.} \textbf{\bibinfo{volume}{121}},
  \bibinfo{pages}{10743} (\bibinfo{year}{1999}).

\bibitem[{\citenamefont{Kodenkandath et~al.}(2001)\citenamefont{Kodenkandath,
  Kumbhar, Zhou, and Wiley}}]{kodenkandath2001}
\bibinfo{author}{\bibfnamefont{T.}~\bibnamefont{Kodenkandath}},
  \bibinfo{author}{\bibfnamefont{A.}~\bibnamefont{Kumbhar}},
  \bibinfo{author}{\bibfnamefont{W.}~\bibnamefont{Zhou}}, \bibnamefont{and}
  \bibinfo{author}{\bibfnamefont{J.}~\bibnamefont{Wiley}},
  \bibinfo{journal}{Inorg. Chem} \textbf{\bibinfo{volume}{40}},
  \bibinfo{pages}{710} (\bibinfo{year}{2001}).

\bibitem[{\citenamefont{Kageyama
  et~al.}(2005{\natexlab{a}})\citenamefont{Kageyama, Kitano, Oba, Nishi, Nagai,
  Hirota, Viciu, Wiley, Yasuda, Baba et~al.}}]{kageyama2005}
\bibinfo{author}{\bibfnamefont{H.}~\bibnamefont{Kageyama}},
  \bibinfo{author}{\bibfnamefont{T.}~\bibnamefont{Kitano}},
  \bibinfo{author}{\bibfnamefont{N.}~\bibnamefont{Oba}},
  \bibinfo{author}{\bibfnamefont{M.}~\bibnamefont{Nishi}},
  \bibinfo{author}{\bibfnamefont{S.}~\bibnamefont{Nagai}},
  \bibinfo{author}{\bibfnamefont{K.}~\bibnamefont{Hirota}},
  \bibinfo{author}{\bibfnamefont{L.}~\bibnamefont{Viciu}},
  \bibinfo{author}{\bibfnamefont{J.~B.} \bibnamefont{Wiley}},
  \bibinfo{author}{\bibfnamefont{J.}~\bibnamefont{Yasuda}},
  \bibinfo{author}{\bibfnamefont{Y.}~\bibnamefont{Baba}},
  \bibinfo{author}{\bibfnamefont{Y.}~\bibnamefont{Ajiro}}, and
  \bibinfo{author}{\bibfnamefont{K.}~\bibnamefont{Yoshimura}},
  \bibinfo{journal}{J. Phys. Soc. Jpn} \textbf{\bibinfo{volume}{74}},
  \bibinfo{pages}{1702} (\bibinfo{year}{2005}{\natexlab{a}}).
  
\bibitem[{\citenamefont{Kageyama
  et~al.}(2005{\natexlab{b}})\citenamefont{Kageyama, Yasuda, Kitano, Totsuka,
  Narumi, Hagiwara, Kindo, Baba, Oba, Ajiro et~al.}}]{kageyama2005-2}
\bibinfo{author}{\bibfnamefont{H.}~\bibnamefont{Kageyama}},
  \bibinfo{author}{\bibfnamefont{J.}~\bibnamefont{Yasuda}},
  \bibinfo{author}{\bibfnamefont{T.}~\bibnamefont{Kitano}},
  \bibinfo{author}{\bibfnamefont{K.}~\bibnamefont{Totsuka}},
  \bibinfo{author}{\bibfnamefont{Y.}~\bibnamefont{Narumi}},
  \bibinfo{author}{\bibfnamefont{M.}~\bibnamefont{Hagiwara}},
  \bibinfo{author}{\bibfnamefont{K.}~\bibnamefont{Kindo}},
  \bibinfo{author}{\bibfnamefont{Y.}~\bibnamefont{Baba}},
  \bibinfo{author}{\bibfnamefont{N.}~\bibnamefont{Oba}},
  \bibinfo{author}{\bibfnamefont{Y.}~\bibnamefont{Ajiro}}, and
  \bibinfo{author}{\bibfnamefont{K.}~\bibnamefont{Yoshimura}},
  \bibinfo{journal}{J. Phys. Soc. Jpn.}
  \textbf{\bibinfo{volume}{74}}, \bibinfo{pages}{3155}
  (\bibinfo{year}{2005}{\natexlab{b}}).
  
\bibitem[{\citenamefont{Caruntu et~al.}(2002)\citenamefont{Caruntu,
  Kodenkandath, and Wiley}}]{caruntu2002}
\bibinfo{author}{\bibfnamefont{G.}~\bibnamefont{Caruntu}},
  \bibinfo{author}{\bibfnamefont{T.~A.} \bibnamefont{Kodenkandath}},
  \bibnamefont{and} \bibinfo{author}{\bibfnamefont{J.~B.} \bibnamefont{Wiley}},
  \bibinfo{journal}{Mater. Res. Bull.} \textbf{\bibinfo{volume}{37}},
  \bibinfo{pages}{593} (\bibinfo{year}{2002}).

\bibitem[{\citenamefont{Yoshida et~al.}(2007)\citenamefont{Yoshida, Ogata,
  Takigawa, Yamaura, Ichihara, Kitano, Kageyama, Ajiro, and
  Yoshimura}}]{yoshida2007}
\bibinfo{author}{\bibfnamefont{M.}~\bibnamefont{Yoshida}},
  \bibinfo{author}{\bibfnamefont{N.}~\bibnamefont{Ogata}},
  \bibinfo{author}{\bibfnamefont{M.}~\bibnamefont{Takigawa}},
  \bibinfo{author}{\bibfnamefont{J.}~\bibnamefont{Yamaura}},
  \bibinfo{author}{\bibfnamefont{M.}~\bibnamefont{Ichihara}},
  \bibinfo{author}{\bibfnamefont{T.}~\bibnamefont{Kitano}},
  \bibinfo{author}{\bibfnamefont{H.}~\bibnamefont{Kageyama}},
  \bibinfo{author}{\bibfnamefont{Y.}~\bibnamefont{Ajiro}}, \bibnamefont{and}
  \bibinfo{author}{\bibfnamefont{K.}~\bibnamefont{Yoshimura}},
  \bibinfo{journal}{J. Phys. Soc. Jpn.} \textbf{\bibinfo{volume}{76}},
  \bibinfo{pages}{104703} (\bibinfo{year}{2007}).

\bibitem[{\citenamefont{Oba et~al.}(2007)\citenamefont{Oba, Kageyama, Saito,
  Azuma, Paulus, Kitano, Ajiro, and Yoshimura}}]{oba2007}
\bibinfo{author}{\bibfnamefont{N.}~\bibnamefont{Oba}},
  \bibinfo{author}{\bibfnamefont{H.}~\bibnamefont{Kageyama}},
  \bibinfo{author}{\bibfnamefont{T.}~\bibnamefont{Saito}},
  \bibinfo{author}{\bibfnamefont{M.}~\bibnamefont{Azuma}},
  \bibinfo{author}{\bibfnamefont{W.}~\bibnamefont{Paulus}},
  \bibinfo{author}{\bibfnamefont{T.}~\bibnamefont{Kitano}},
  \bibinfo{author}{\bibfnamefont{Y.}~\bibnamefont{Ajiro}}, \bibnamefont{and}
  \bibinfo{author}{\bibfnamefont{K.}~\bibnamefont{Yoshimura}},
  \bibinfo{journal}{J. Magn. Magn. Mater.} \textbf{\bibinfo{volume}{310}},
  \bibinfo{pages}{1337} (\bibinfo{year}{2007}); cond-mat/0511744.

\bibitem[{foo({\natexlab{a}})}]{foot1}
\bibinfo{note}{Basically, one can overcome multiple scattering by using the
  precession technique, and high-resolution electron microscopy provides
  additional structural information on the local scale. In general, the
  structure solution and refinement from electron diffraction and electron
  microscopy data are possible (see Ref.~\onlinecite{elcryst}), but they are
  still far from being routine procedures in modern structure analysis.}

\bibitem[{\citenamefont{Kitada et~al.}(2007)\citenamefont{Kitada, Hiroi,
  Tsujimoto, Kitano, Kageyama, Ajiro, and Yoshimura}}]{kitada2007}
\bibinfo{author}{\bibfnamefont{A.}~\bibnamefont{Kitada}},
  \bibinfo{author}{\bibfnamefont{Z.}~\bibnamefont{Hiroi}},
  \bibinfo{author}{\bibfnamefont{Y.}~\bibnamefont{Tsujimoto}},
  \bibinfo{author}{\bibfnamefont{T.}~\bibnamefont{Kitano}},
  \bibinfo{author}{\bibfnamefont{H.}~\bibnamefont{Kageyama}},
  \bibinfo{author}{\bibfnamefont{Y.}~\bibnamefont{Ajiro}}, \bibnamefont{and}
  \bibinfo{author}{\bibfnamefont{K.}~\bibnamefont{Yoshimura}},
  \bibinfo{journal}{J. Phys. Soc. Jpn.} \textbf{\bibinfo{volume}{76}},
  \bibinfo{pages}{093706} (\bibinfo{year}{2007}).

\bibitem[{\citenamefont{Whangbo and Dai}(2006)}]{whangbo2006}
\bibinfo{author}{\bibfnamefont{M.-H.} \bibnamefont{Whangbo}} \bibnamefont{and}
  \bibinfo{author}{\bibfnamefont{D.}~\bibnamefont{Dai}},
  \bibinfo{journal}{Inorg. Chem.} \textbf{\bibinfo{volume}{45}},
  \bibinfo{pages}{6227} (\bibinfo{year}{2006}).

\bibitem[{\citenamefont{Rosner and Pickett}(2003)}]{lib}
\bibinfo{author}{\bibfnamefont{H.}~\bibnamefont{Rosner}} \bibnamefont{and}
  \bibinfo{author}{\bibfnamefont{W.~E.} \bibnamefont{Pickett}},
  \bibinfo{journal}{Phys. Rev. B} \textbf{\bibinfo{volume}{67}},
  \bibinfo{pages}{054104} (\bibinfo{year}{2003}).

\bibitem[{\citenamefont{Liechtenstein et~al.}(1995)\citenamefont{Liechtenstein,
  Anisimov, and Zaanen}}]{kcuf3}
\bibinfo{author}{\bibfnamefont{A.~I.} \bibnamefont{Liechtenstein}},
  \bibinfo{author}{\bibfnamefont{V.~I.} \bibnamefont{Anisimov}},
  \bibnamefont{and} \bibinfo{author}{\bibfnamefont{J.}~\bibnamefont{Zaanen}},
  \bibinfo{journal}{Phys. Rev. B} \textbf{\bibinfo{volume}{52}},
  \bibinfo{pages}{R5467} (\bibinfo{year}{1995}).

\bibitem[{\citenamefont{Leonov et~al.}(2008)\citenamefont{Leonov, Binggeli,
  Korotin, Anisimov, Stoji\'c, and Vollhardt}}]{kcuf3-2}
\bibinfo{author}{\bibfnamefont{I.}~\bibnamefont{Leonov}},
  \bibinfo{author}{\bibfnamefont{N.}~\bibnamefont{Binggeli}},
  \bibinfo{author}{\bibfnamefont{Dm.}~\bibnamefont{Korotin}},
  \bibinfo{author}{\bibfnamefont{V.~I.} \bibnamefont{Anisimov}},
  \bibinfo{author}{\bibfnamefont{N.}~\bibnamefont{Stoji\'c}}, \bibnamefont{and}
  \bibinfo{author}{\bibfnamefont{D.}~\bibnamefont{Vollhardt}},
  \bibinfo{journal}{Phys. Rev. Lett.} \textbf{\bibinfo{volume}{101}},
  \bibinfo{pages}{096405} (\bibinfo{year}{2008}); arXiv:0804.1093.

\bibitem[{\citenamefont{Kasinathan et~al.}(2007)\citenamefont{Kasinathan,
  Koepernik, Nitzsche, and Rosner}}]{cs2agf4}
\bibinfo{author}{\bibfnamefont{D.}~\bibnamefont{Kasinathan}},
  \bibinfo{author}{\bibfnamefont{K.}~\bibnamefont{Koepernik}},
  \bibinfo{author}{\bibfnamefont{U.}~\bibnamefont{Nitzsche}}, \bibnamefont{and}
  \bibinfo{author}{\bibfnamefont{H.}~\bibnamefont{Rosner}},
  \bibinfo{journal}{Phys. Rev. Lett.} \textbf{\bibinfo{volume}{99}},
  \bibinfo{pages}{247210} (\bibinfo{year}{2007}).

\bibitem[{\citenamefont{Wang et~al.}(2007)\citenamefont{Wang, Guo, and
  He}}]{tbmn2o5-computation}
\bibinfo{author}{\bibfnamefont{C.}~\bibnamefont{Wang}},
  \bibinfo{author}{\bibfnamefont{G.-C.} \bibnamefont{Guo}}, \bibnamefont{and}
  \bibinfo{author}{\bibfnamefont{L.}~\bibnamefont{He}}, \bibinfo{journal}{Phys.
  Rev. Lett.} \textbf{\bibinfo{volume}{99}}, \bibinfo{pages}{177202}
  (\bibinfo{year}{2007}); arXiv:0711.2539.

\bibitem[{\citenamefont{Picozzi et~al.}(2007)\citenamefont{Picozzi, Yamauchi,
  Sanyal, Sergienko, and Dagotto}}]{homno3}
\bibinfo{author}{\bibfnamefont{S.}~\bibnamefont{Picozzi}},
  \bibinfo{author}{\bibfnamefont{K.}~\bibnamefont{Yamauchi}},
  \bibinfo{author}{\bibfnamefont{B.}~\bibnamefont{Sanyal}},
  \bibinfo{author}{\bibfnamefont{I.~A.} \bibnamefont{Sergienko}},
  \bibnamefont{and} \bibinfo{author}{\bibfnamefont{E.}~\bibnamefont{Dagotto}},
  \bibinfo{journal}{Phys. Rev. Lett.} \textbf{\bibinfo{volume}{99}},
  \bibinfo{pages}{227201} (\bibinfo{year}{2007}); arXiv:0704.3578.

\bibitem[{\citenamefont{Koo et~al.}(2007)\citenamefont{Koo, Song, Ji, Lee,
  Park, Jang, Yang, Park, Jeong, Lee et~al.}}]{tbmn2o5-exp}
\bibinfo{author}{\bibfnamefont{J.}~\bibnamefont{Koo}},
  \bibinfo{author}{\bibfnamefont{C.}~\bibnamefont{Song}},
  \bibinfo{author}{\bibfnamefont{S.}~\bibnamefont{Ji}},
  \bibinfo{author}{\bibfnamefont{J.-S.} \bibnamefont{Lee}},
  \bibinfo{author}{\bibfnamefont{J.}~\bibnamefont{Park}},
  \bibinfo{author}{\bibfnamefont{T.-H.} \bibnamefont{Jang}},
  \bibinfo{author}{\bibfnamefont{C.-H.} \bibnamefont{Yang}},
  \bibinfo{author}{\bibfnamefont{J.-H.} \bibnamefont{Park}},
  \bibinfo{author}{\bibfnamefont{Y.~H.} \bibnamefont{Jeong}},
  \bibinfo{author}{\bibfnamefont{K.-B.} \bibnamefont{Lee}},
  \bibnamefont{et~al.}, \bibinfo{journal}{Phys. Rev. Lett.}
  \textbf{\bibinfo{volume}{99}}, \bibinfo{pages}{197601}
  (\bibinfo{year}{2007}); arXiv:0704.0533.

\bibitem[{\citenamefont{Koepernik and Eschrig}(1999)}]{fplo}
\bibinfo{author}{\bibfnamefont{K.}~\bibnamefont{Koepernik}} \bibnamefont{and}
  \bibinfo{author}{\bibfnamefont{H.}~\bibnamefont{Eschrig}},
  \bibinfo{journal}{Phys. Rev. B} \textbf{\bibinfo{volume}{59}},
  \bibinfo{pages}{1743} (\bibinfo{year}{1999}).

\bibitem[{\citenamefont{Perdew and Wang}(1992)}]{perdew}
\bibinfo{author}{\bibfnamefont{J.~P.} \bibnamefont{Perdew}} \bibnamefont{and}
  \bibinfo{author}{\bibfnamefont{Y.}~\bibnamefont{Wang}},
  \bibinfo{journal}{Phys. Rev. B} \textbf{\bibinfo{volume}{45}},
  \bibinfo{pages}{13244} (\bibinfo{year}{1992}).

\bibitem[{\citenamefont{Marzari and Vanderbilt}(1997)}]{wannier}
\bibinfo{author}{\bibfnamefont{N.}~\bibnamefont{Marzari}} \bibnamefont{and}
  \bibinfo{author}{\bibfnamefont{D.}~\bibnamefont{Vanderbilt}},
  \bibinfo{journal}{Phys. Rev.~B} \textbf{\bibinfo{volume}{56}},
  \bibinfo{pages}{12847} (\bibinfo{year}{1997}); cond-mat/9707145.

\bibitem[{foo({\natexlab{b}})}]{foot2}
\bibinfo{note}{We should emphasize that there are several extensively studied
  layered Cl-containing copper oxides (e.g., Sr$_2$CuO$_2$Cl$_2$). These
  compounds can be considered as oxychlorides from chemical point of view, but
  their crystal and electronic structures are similar to that of conventional
  Cu$^{+2}$-containing oxides, because Cu atoms are situated in CuO$_4$
  plaquettes, while Cl atoms form long axial Cu--Cl bonds. As a result, the Cl
  orbitals do not overlap with the half-filled Cu $d_{x^2-y^2}$ orbital and
  bear little influence on the physics. Electronic structures of "true" copper
  oxyhalides (with halogen atoms involved in the copper plaquettes) remain
  basically unexplored. To the best of our knowledge, band structure
  calculations are reported for five compounds only: Cu$_2$Te$_2$O$_5$X$_2$ (X
  = Cl, Br),\cite{cute2252} Cu$_4$Te$_5$O$_{12}$Cl$_4$
  (Ref.~\onlinecite{cute45124}), and, quite recently, CuCl$_2$ along with
  CuCl$_2\cdot 2$H$_2$O.\cite{cucl2} The on-site Coulomb repulsion
  parameters are discussed for the latter case only, and the values of
  $U_{\eff}=4$~eV, $U_d=6.0-8.5$~eV are used.}

\bibitem[{\citenamefont{Schmitt et~al.}()\citenamefont{Schmitt, Janson,
  Schmidt, Scnhelle, Drechsler, and Rosner}}]{cucl2}
\bibinfo{author}{\bibfnamefont{M.}~\bibnamefont{Schmitt}},
  \bibinfo{author}{\bibfnamefont{O.}~\bibnamefont{Janson}},
  \bibinfo{author}{\bibfnamefont{M.}~\bibnamefont{Schmidt}},
  \bibinfo{author}{\bibfnamefont{W.}~\bibnamefont{Scnhelle}},
  \bibinfo{author}{\bibfnamefont{S.-L.} \bibnamefont{Drechsler}},
  \bibnamefont{and} \bibinfo{author}{\bibfnamefont{H.}~\bibnamefont{Rosner}},
  \bibinfo{note}{Phys. Rev. B (in press)}; arXiv:0905.4038.

\bibitem[{\citenamefont{Johannes et~al.}(2006)\citenamefont{Johannes, Richter,
  Drechsler, and Rosner}}]{sr2cup2o8}
\bibinfo{author}{\bibfnamefont{M.~D.} \bibnamefont{Johannes}},
  \bibinfo{author}{\bibfnamefont{J.}~\bibnamefont{Richter}},
  \bibinfo{author}{\bibfnamefont{S.-L.} \bibnamefont{Drechsler}},
  \bibnamefont{and} \bibinfo{author}{\bibfnamefont{H.}~\bibnamefont{Rosner}},
  \bibinfo{journal}{Phys. Rev. B} \textbf{\bibinfo{volume}{74}},
  \bibinfo{pages}{174435} (\bibinfo{year}{2006}); cond-mat/0609430.

\bibitem[{\citenamefont{Janson et~al.}(2007)\citenamefont{Janson, Kuzian,
  Drechsler, and Rosner}}]{bi2cuo4}
\bibinfo{author}{\bibfnamefont{O.}~\bibnamefont{Janson}},
  \bibinfo{author}{\bibfnamefont{R.~O.} \bibnamefont{Kuzian}},
  \bibinfo{author}{\bibfnamefont{S.-L.} \bibnamefont{Drechsler}},
  \bibnamefont{and} \bibinfo{author}{\bibfnamefont{H.}~\bibnamefont{Rosner}},
  \bibinfo{journal}{Phys. Rev. B} \textbf{\bibinfo{volume}{76}},
  \bibinfo{pages}{115119} (\bibinfo{year}{2007}).

\bibitem[{\citenamefont{Enderle et~al.}(2005)\citenamefont{Enderle, Mukherjee,
  F{\r a}k, Kremer, Broto, Rosner, Drechsler, Richter, Malek, Prokofiev
  et~al.}}]{licuvo4}
\bibinfo{author}{\bibfnamefont{M.}~\bibnamefont{Enderle}},
  \bibinfo{author}{\bibfnamefont{C.}~\bibnamefont{Mukherjee}},
  \bibinfo{author}{\bibfnamefont{B.}~\bibnamefont{F{\r a}k}},
  \bibinfo{author}{\bibfnamefont{R.~K.} \bibnamefont{Kremer}},
  \bibinfo{author}{\bibfnamefont{J.-M.} \bibnamefont{Broto}},
  \bibinfo{author}{\bibfnamefont{H.}~\bibnamefont{Rosner}},
  \bibinfo{author}{\bibfnamefont{S.-L.} \bibnamefont{Drechsler}},
  \bibinfo{author}{\bibfnamefont{J.}~\bibnamefont{Richter}},
  \bibinfo{author}{\bibfnamefont{J.}~\bibnamefont{Malek}},
  \bibinfo{author}{\bibfnamefont{A.}~\bibnamefont{Prokofiev}},
  \bibinfo{author}{\bibfnamefont{W.}~\bibnamefont{Assmus}},
  \bibinfo{author}{\bibfnamefont{S.}~\bibnamefont{Pujol}},
  \bibinfo{author}{\bibfnamefont{J.-L.}~\bibnamefont{Raggazoni}},
  \bibinfo{author}{\bibfnamefont{H.}~\bibnamefont{Rakoto}},
  \bibinfo{author}{\bibfnamefont{M.}~\bibnamefont{Rheinst{\"a}dter}}, and
  \bibinfo{author}{\bibfnamefont{H.~M.}~\bibnamefont{R\o nnow}},
  \bibinfo{journal}{Europhys. Lett.}
  \textbf{\bibinfo{volume}{70}}, \bibinfo{pages}{237} (\bibinfo{year}{2005}).
  
\bibitem[{\citenamefont{Kasinathan et~al.}(2008)\citenamefont{Kasinathan,
  Koepernik, and Rosner}}]{cusb2o6}
\bibinfo{author}{\bibfnamefont{D.}~\bibnamefont{Kasinathan}},
  \bibinfo{author}{\bibfnamefont{K.}~\bibnamefont{Koepernik}},
  \bibnamefont{and} \bibinfo{author}{\bibfnamefont{H.}~\bibnamefont{Rosner}},
  \bibinfo{journal}{Phys. Rev. Lett.} \textbf{\bibinfo{volume}{100}},
  \bibinfo{pages}{237202} (\bibinfo{year}{2008}); arXiv:0805.4080.

\bibitem[{foo({\natexlab{c}})}]{foot4}
\bibinfo{note}{One can get a feeling of these values by considering the
  LSDA+$U$ calculations for conventional cuprates with CuO$_4$ plaquettes.
  Depending on the computational method and on the specific compound, the $U_d$
  values in the range $6.5-8$ eV are used for the evaluation of the exchange
  couplings (see, e.g., Refs.~\onlinecite{sr2cup2o8} and \onlinecite{bi2cuo4}).
  The $3p$ orbitals of chlorine are spatially more extended as compared to the
  $2p$ orbitals of oxygen. Therefore, one may expect more effective screening
  (hence, smaller on-site repulsion) in copper oxychlorides. Based on these
  considerations, we employ the "cuprate-like" $U_d$ of 7.5 eV along with the
  smaller $U_d$ values in our LSDA+$U$ calculations.}

\bibitem[{\citenamefont{Valenti et~al.}(2003)\citenamefont{Valenti,
  Saha-Dasgupta, Gros, and Rosner}}]{cute2252}
\bibinfo{author}{\bibfnamefont{R.}~\bibnamefont{Valenti}},
  \bibinfo{author}{\bibfnamefont{T.}~\bibnamefont{Saha-Dasgupta}},
  \bibinfo{author}{\bibfnamefont{C.}~\bibnamefont{Gros}}, \bibnamefont{and}
  \bibinfo{author}{\bibfnamefont{H.}~\bibnamefont{Rosner}},
  \bibinfo{journal}{Phys. Rev. B} \textbf{\bibinfo{volume}{67}},
  \bibinfo{pages}{245110} (\bibinfo{year}{2003}); cond-mat/0301119.

\bibitem[{\citenamefont{Rahaman et~al.}(2007)\citenamefont{Rahaman, Jeschke,
  Valenti, and Saha-Dasgupta}}]{cute45124}
\bibinfo{author}{\bibfnamefont{B.}~\bibnamefont{Rahaman}},
  \bibinfo{author}{\bibfnamefont{H.~O.} \bibnamefont{Jeschke}},
  \bibinfo{author}{\bibfnamefont{R.}~\bibnamefont{Valenti}}, \bibnamefont{and}
  \bibinfo{author}{\bibfnamefont{T.}~\bibnamefont{Saha-Dasgupta}},
  \bibinfo{journal}{Phys. Rev. B} \textbf{\bibinfo{volume}{75}},
  \bibinfo{pages}{024404} (\bibinfo{year}{2007}); cond-mat/0608598.

\bibitem[{\citenamefont{Schmidt et~al.}(2008)\citenamefont{Schmidt, Albrecht,
  Wippermann, Blankenburg, Rauls, Fuchs, R\"odl, Furthm\"uller, and
  Hermann}}]{linbo3}
\bibinfo{author}{\bibfnamefont{W.~G.} \bibnamefont{Schmidt}},
  \bibinfo{author}{\bibfnamefont{M.}~\bibnamefont{Albrecht}},
  \bibinfo{author}{\bibfnamefont{S.}~\bibnamefont{Wippermann}},
  \bibinfo{author}{\bibfnamefont{S.}~\bibnamefont{Blankenburg}},
  \bibinfo{author}{\bibfnamefont{E.}~\bibnamefont{Rauls}},
  \bibinfo{author}{\bibfnamefont{F.}~\bibnamefont{Fuchs}},
  \bibinfo{author}{\bibfnamefont{C.}~\bibnamefont{R\"odl}},
  \bibinfo{author}{\bibfnamefont{J.}~\bibnamefont{Furthm\"uller}},
  \bibnamefont{and} \bibinfo{author}{\bibfnamefont{A.}~\bibnamefont{Hermann}},
  \bibinfo{journal}{Phys. Rev. B} \textbf{\bibinfo{volume}{77}},
  \bibinfo{pages}{035106} (\bibinfo{year}{2008}).

\bibitem[{\citenamefont{Rosner et~al.}(2003)\citenamefont{Rosner, Singh, Zheng,
  Oitmaa, and Pickett}}]{rosner2003}
\bibinfo{author}{\bibfnamefont{H.}~\bibnamefont{Rosner}},
  \bibinfo{author}{\bibfnamefont{R.~R.~P.} \bibnamefont{Singh}},
  \bibinfo{author}{\bibfnamefont{W.~H.} \bibnamefont{Zheng}},
  \bibinfo{author}{\bibfnamefont{J.}~\bibnamefont{Oitmaa}}, \bibnamefont{and}
  \bibinfo{author}{\bibfnamefont{W.~E.} \bibnamefont{Pickett}},
  \bibinfo{journal}{Phys. Rev. B} \textbf{\bibinfo{volume}{67}},
  \bibinfo{pages}{014416} (\bibinfo{year}{2003}).

\bibitem[{\citenamefont{Schmidt et~al.}(2007)\citenamefont{Schmidt, Thalmeier,
  and Shannon}}]{schmidt2007}
\bibinfo{author}{\bibfnamefont{B.}~\bibnamefont{Schmidt}},
  \bibinfo{author}{\bibfnamefont{P.}~\bibnamefont{Thalmeier}},
  \bibnamefont{and} \bibinfo{author}{\bibfnamefont{N.}~\bibnamefont{Shannon}},
  \bibinfo{journal}{Phys. Rev. B} \textbf{\bibinfo{volume}{76}},
  \bibinfo{pages}{125113} (\bibinfo{year}{2007}); arXiv:0705.3094.

\bibitem[{\citenamefont{Woodward}(1997)}]{woodward}
\bibinfo{author}{\bibfnamefont{P.~M.} \bibnamefont{Woodward}},
  \bibinfo{journal}{Acta Cryst.} \textbf{\bibinfo{volume}{B53}},
  \bibinfo{pages}{44} (\bibinfo{year}{1997}).

\bibitem[{foo({\natexlab{d}})}]{foot5}
\bibinfo{note}{The stability of the perovskite structure can be evaluated by
  calculating a simple geometrical tolerance factor ($t_f$). For the ideal
  (cubic) perovskite structure of the ABO$_3$ compound,
  $t_f=(r_{\text{A}}+r_{\text{O}})/\sqrt 2(r_{\text{B}}+r_{\text{O}})$ amounts
  to 1. In case of the [LaNb$_2$O$_7$] block, $t_f\simeq 0.94$ implying that
  the La atom is a bit too small for the framework of the NbO$_6$ octahedra.
  Then the small size of the La atom may be tolerated by the tiltings of the
  NbO$_6$ octahedra, since such tiltings lead to the reduction in La--O
  distances, see Ref.~\onlinecite{woodward} for details.}

\bibitem[{\citenamefont{Pickett}(1997)}]{pickett}
\bibinfo{author}{\bibfnamefont{W.~E.} \bibnamefont{Pickett}},
  \bibinfo{journal}{Phys. Rev. Lett.} \textbf{\bibinfo{volume}{79}},
  \bibinfo{pages}{1746} (\bibinfo{year}{1997}); cond-mat/9704203.

\bibitem[{\citenamefont{Korotin et~al.}(1999)\citenamefont{Korotin, Elfimov,
  Anisimov, Troyer, and Khomskii}}]{korotin}
\bibinfo{author}{\bibfnamefont{M.~A.} \bibnamefont{Korotin}},
  \bibinfo{author}{\bibfnamefont{I.~S.} \bibnamefont{Elfimov}},
  \bibinfo{author}{\bibfnamefont{V.~I.} \bibnamefont{Anisimov}},
  \bibinfo{author}{\bibfnamefont{M.}~\bibnamefont{Troyer}}, \bibnamefont{and}
  \bibinfo{author}{\bibfnamefont{D.~I.} \bibnamefont{Khomskii}},
  \bibinfo{journal}{Phys. Rev. Lett.} \textbf{\bibinfo{volume}{83}},
  \bibinfo{pages}{1387} (\bibinfo{year}{1999}); cond-mat/9901214.

\bibitem[{\citenamefont{Oba et~al.}(2006)\citenamefont{Oba, Kageyama, Kitano,
  Yasuda, Baba, Nishi, Hirota, Narumi, Hagiwara, Kindo et~al.}}]{oba2006}
\bibinfo{author}{\bibfnamefont{N.}~\bibnamefont{Oba}},
  \bibinfo{author}{\bibfnamefont{H.}~\bibnamefont{Kageyama}},
  \bibinfo{author}{\bibfnamefont{T.}~\bibnamefont{Kitano}},
  \bibinfo{author}{\bibfnamefont{J.}~\bibnamefont{Yasuda}},
  \bibinfo{author}{\bibfnamefont{Y.}~\bibnamefont{Baba}},
  \bibinfo{author}{\bibfnamefont{M.}~\bibnamefont{Nishi}},
  \bibinfo{author}{\bibfnamefont{K.}~\bibnamefont{Hirota}},
  \bibinfo{author}{\bibfnamefont{Y.}~\bibnamefont{Narumi}},
  \bibinfo{author}{\bibfnamefont{M.}~\bibnamefont{Hagiwara}},
  \bibinfo{author}{\bibfnamefont{K.}~\bibnamefont{Kindo}},
  \bibinfo{author}{\bibfnamefont{T.}~\bibnamefont{Saito}},
  \bibinfo{author}{\bibfnamefont{T.}~\bibnamefont{Ajiro}}, and
  \bibinfo{author}{\bibfnamefont{K.}~\bibnamefont{Yoshimura}},
  \bibinfo{journal}{J. Phys. Soc. Jpn}
  \textbf{\bibinfo{volume}{75}}, \bibinfo{pages}{113601}
  (\bibinfo{year}{2006}).

\bibitem[{\citenamefont{Tsujimoto et~al.}(2007)\citenamefont{Tsujimoto, Baba,
  Oba, Kageyama, Fukui, Narumi, Kindo, Saito, Takano, Ajiro
  et~al.}}]{tsujimoto2007}
\bibinfo{author}{\bibfnamefont{Y.}~\bibnamefont{Tsujimoto}},
  \bibinfo{author}{\bibfnamefont{Y.}~\bibnamefont{Baba}},
  \bibinfo{author}{\bibfnamefont{N.}~\bibnamefont{Oba}},
  \bibinfo{author}{\bibfnamefont{H.}~\bibnamefont{Kageyama}},
  \bibinfo{author}{\bibfnamefont{T.}~\bibnamefont{Fukui}},
  \bibinfo{author}{\bibfnamefont{Y.}~\bibnamefont{Narumi}},
  \bibinfo{author}{\bibfnamefont{K.}~\bibnamefont{Kindo}},
  \bibinfo{author}{\bibfnamefont{T.}~\bibnamefont{Saito}},
  \bibinfo{author}{\bibfnamefont{M.}~\bibnamefont{Takano}},
  \bibinfo{author}{\bibfnamefont{Y.}~\bibnamefont{Ajiro}}, and
  \bibinfo{author}{\bibfnamefont{K.}~\bibnamefont{Yoshimura}},
  \bibinfo{journal}{J. Phys. Soc. Jpn.}
  \textbf{\bibinfo{volume}{76}}, \bibinfo{pages}{063711}
  (\bibinfo{year}{2007}).

\bibitem[{\citenamefont{Tsujimoto et~al.}(2008)\citenamefont{Tsujimoto,
  Kageyama, Baba, Kitada, Yamamoto, Narumi, Kindo, Nishi, Carlo, Aczel
  et~al.}}]{tsujimoto2008}
\bibinfo{author}{\bibfnamefont{Y.}~\bibnamefont{Tsujimoto}},
  \bibinfo{author}{\bibfnamefont{H.}~\bibnamefont{Kageyama}},
  \bibinfo{author}{\bibfnamefont{Y.}~\bibnamefont{Baba}},
  \bibinfo{author}{\bibfnamefont{A.}~\bibnamefont{Kitada}},
  \bibinfo{author}{\bibfnamefont{T.}~\bibnamefont{Yamamoto}},
  \bibinfo{author}{\bibfnamefont{Y.}~\bibnamefont{Narumi}},
  \bibinfo{author}{\bibfnamefont{K.}~\bibnamefont{Kindo}},
  \bibinfo{author}{\bibfnamefont{M.}~\bibnamefont{Nishi}},
  \bibinfo{author}{\bibfnamefont{J.~P.} \bibnamefont{Carlo}},
  \bibinfo{author}{\bibfnamefont{A.~A.} \bibnamefont{Aczel}},
  \bibinfo{author}{\bibfnamefont{T.~J.} \bibnamefont{Williams}},
  \bibinfo{author}{\bibfnamefont{T.} \bibnamefont{Goko}},
  \bibinfo{author}{\bibfnamefont{G.~M.} \bibnamefont{Luke}},
  \bibinfo{author}{\bibfnamefont{Y.~J.} \bibnamefont{Uemura}},
  \bibinfo{author}{\bibfnamefont{Y.} \bibnamefont{Ueda}},
  \bibinfo{author}{\bibfnamefont{Y.} \bibnamefont{Ajiro}}, and
  \bibinfo{author}{\bibfnamefont{K.} \bibnamefont{Yoshimura}},
  \bibinfo{journal}{Phys. Rev. B}
  \textbf{\bibinfo{volume}{78}}, \bibinfo{pages}{214410}
  (\bibinfo{year}{2008}).
  
\bibitem[{\citenamefont{Yoshida et~al.}(2008)\citenamefont{Yoshida, Ogata,
  Takigawa, Kitano, Kageyama, Ajiro, and Yoshimura}}]{yoshida2008}
\bibinfo{author}{\bibfnamefont{M.}~\bibnamefont{Yoshida}},
  \bibinfo{author}{\bibfnamefont{N.}~\bibnamefont{Ogata}},
  \bibinfo{author}{\bibfnamefont{M.}~\bibnamefont{Takigawa}},
  \bibinfo{author}{\bibfnamefont{T.}~\bibnamefont{Kitano}},
  \bibinfo{author}{\bibfnamefont{H.}~\bibnamefont{Kageyama}},
  \bibinfo{author}{\bibfnamefont{Y.}~\bibnamefont{Ajiro}}, \bibnamefont{and}
  \bibinfo{author}{\bibfnamefont{K.}~\bibnamefont{Yoshimura}},
  \bibinfo{journal}{J. Phys. Soc. Jpn.} \textbf{\bibinfo{volume}{77}},
  \bibinfo{pages}{104705} (\bibinfo{year}{2008}).

\bibitem[{\citenamefont{Zou and Hovm\"oller}(2008)}]{elcryst}
\bibinfo{author}{\bibfnamefont{X.~D.} \bibnamefont{Zou}} \bibnamefont{and}
  \bibinfo{author}{\bibfnamefont{S.}~\bibnamefont{Hovm\"oller}},
  \bibinfo{journal}{Acta Cryst.} \textbf{\bibinfo{volume}{A64}},
  \bibinfo{pages}{149} (\bibinfo{year}{2008}).

\end{thebibliography}
\end{document}